\documentclass[12pt]{article} 
\pdfoutput=1
\usepackage{amssymb,graphicx}
\usepackage{epstopdf}
\usepackage{amsmath,amsfonts}
\usepackage{epsfig} 
\usepackage{graphicx,graphics}

\usepackage{xcolor}
 \usepackage[unicode=true,pdfusetitle,
 bookmarks=true,bookmarksnumbered=true,bookmarksopen=true,bookmarksopenlevel=3,
 breaklinks=false,pdfborder={0 0 1},backref=false,colorlinks=true,citecolor=red]
 {hyperref}

\begin{document}

\sloppy

\title{\bf Heating up Peccei--Quinn scale}
\author{Sabir Ramazanov and Rome Samanta\\
\small{\em CEICO, FZU-Institute of Physics of the Czech Academy of Sciences,}\\
\small{\em Na Slovance 1999/2, 182 00 Prague 8, Czech Republic}\\
}
 
{\let\newpage\relax\maketitle}

\begin{abstract}

We discuss production of QCD axion dark matter in a novel scenario, which assumes time-varying scale of Peccei-Quinn symmetry breaking. The latter decreases as the Universe's temperature at early times and eventually stabilises at a large constant value. Such behavior is caused by the portal interaction between the complex field carrying Peccei-Quinn charge and a Higgs-like scalar, which is in thermal equilibrium with primordial plasma. In this scenario, axions are efficiently produced during the parametric resonance decay of the 
complex Peccei-Quinn field, relaxing to the minimum of its potential in the radiation-dominated stage. Notably, this process is not affected by the Universe's expansion rate and allows to generate the required abundance of dark matter independently of an axion mass. Phenomenological constraints on the model parameter space depend 
on the number density of radial field fluctuations, which are also generically excited along with axions, and the rate of their thermalization in the primordial plasma. 
For the ratio of radial field and axion particles number densities larger than $\sim 0.01$ at the end of parametric resonance decay, the combination of cosmological and astrophysical observations with the CAST limit confines 
the Peccei-Quinn scale to a narrow range of values $\sim 10^{8}~\mbox{GeV}$, --- 
this paves the way for ruling out our scenario with the near future searches for axions.

\end{abstract}

\section{Introduction} 
Axions, pseudo-Nambu-Goldstone bosons of spontaneously broken Peccei-Quinn symmetry~\cite{Peccei:1977hh, Peccei:1977ur, Wilczek:1977pj, Weinberg:1977ma}, provide an elegant resolution of the strong CP problem in Quantum Chromodynamics (QCD). Instanton effects endow an axion with a mass $m_a$, inversely proportional to the scale of Peccei-Quinn symmetry breaking $f_{PQ}$. The latter must be set well above the Higgs field expectation value $v_{SM}$ to warrant small axion couplings to Standard Model (SM) particles. There are two benchmark axion setups realising the hierarchy $f_{PQ} \gg v_{SM}$: the Kim-Shifman-Veinshtein-Zakharov (KSVZ) scenario and its variations~\cite{Kim:1979if, Shifman:1979if}, plus the Dine-Fischler-Srednicki-Zhitnitski (DFSZ) type of models~\cite{Dine:1981rt, Zhitnitsky:1980tq}. For substantially large $f_{PQ}$, axions are stable on 
cosmological time scales, and thus can be regarded as promising dark matter candidates, if produced abundantly in the early Universe~\cite{Preskill:1982cy, Abbott:1982af, Dine:1982ah}. In the present work, we are primarily interested 
in the case of QCD dark matter axions.

While very weakly coupled to matter fields, axions are actively searched for in astrophysical and cosmological backgrounds, as well as with Earth-based facilities~\cite{DiLuzio:2020wdo}. 
The search programs strongly depend on the range of values of $f_{PQ}$, which in turn is limited by the choice of axion production mechanisms. 
Most commonly, axions are produced through vacuum realignment~\cite{Preskill:1982cy, Abbott:1982af, Dine:1982ah} or decay of topological defects~\cite{Davis:1986xc, Kawasaki:2014sqa} in the range\footnote{For simplicity, in this work
 we identify the axion decay constant $f_a$ and Peccei-Quinn symmetry breaking scale $f_{PQ}$. See the remark in the beginning of Section~\ref{sec:darkmatter}.} $10^{11}~\mbox{GeV} \lesssim f_{PQ} \lesssim 10^{12}~\mbox{GeV}$. For such large values of $f_{PQ}$, axions are unlikely to have a sizeable impact on stellar evolution. In particular, their effect on the duration of neutrino burst from the Supernova 1987A suggests the rough limit $f_{PQ} \gtrsim 4 \times 10^{8}~\mbox{GeV}$~\cite{Raffelt:2006cw}. Furthermore, anomalous cooling of horizontal branch stars favours relatively small $f_{PQ} \sim 10^{8}~\mbox{GeV}$ corresponding to axion masses $m_a \sim 5 \cdot 10^{-2}~\mbox{eV}$~\cite{Giannotti:2015kwo, Giannotti:2017hny}. These considerations motivate devising new mechanisms capable of generating axions in the range of parameters interesting for astrophysics, cf.~Refs.~\cite{Co:2017mop, Co:2020dya, Harigaya:2019qnl, Nakayama:2021avl, Co:2019jts, Hook:2019hdk}.

We propose a novel scenario of QCD axion genesis at high temperatures, which leads to the correct dark matter abundance for the masses $m_a \gtrsim 10^{-2}~\mbox{eV}$. 
We assume that the scale $f_{PQ}$ decreases with the Universe's temperature $T$ at sufficiently early times, i.e., $f_{PQ} (t) \propto T(t)$, and later on, stabilises at a constant value $f^{0}_{PQ}$. 
This is achieved by the portal coupling of the complex field $S$ carrying Peccei-Quinn charge to a thermal scalar $\phi$ in the primordial plasma; see Section~\ref{sec:themodel}. The scale $f^{0}_{PQ}$ is related to variance of the field $\phi$ by 
$f_{PQ} \propto \langle \phi^{\dagger} \phi \rangle^{1/2}$. Thus, if the field $\phi$ is in the spontaneously broken phase at low temperatures with the expectation value $v_{\phi}$, one has $f^{0}_{PQ} \propto v_{\phi}$. 
At very early times $\langle \phi^{\dagger} \phi \rangle^{1/2} \sim T$, and we achieve the desired temperature dependence of the scale $f_{PQ}$. (See also Ref.~\cite{Allali:2022yvx} discussing different forms of 
Peccei-Quinn scale's temporal variation with a similar goal of altering axion parameter space.). We would like to remark that the generic idea of spontaneous symmetry breaking caused by thermal fluctuations of hot primordial plasma is rather old~\cite{Weinberg:1974hy}. 
It was later implemented in various contexts: topological defects~\cite{Vilenkin:1981zs, Emond:2021vts, Babichev:2021uvl}, baryon asymmetry generation~\cite{Dodelson:1989cq}, dark matter production~\cite{Ramazanov:2021eya}, and conformal field 
theories~\cite{Chai:2020onq}. The present work continues this research line by applying the idea to axion models.

 In our scenario, axions appear through parametric resonance when the radial part of the Peccei-Quinn complex scalar $|S|$ relaxes to the minimum of its spontaneously broken potential (Section~\ref{sec:resonance}). Compared to Refs.~\cite{Co:2017mop, Co:2020dya, Ema:2017krp}, we consider the situation when the initial amplitude of oscillations is smaller than $f_{PQ}$. This is a natural option in our case, given huge values of $f_{PQ}$ at large temperatures and initial conditions for the field $S$ set by cosmic inflation. As a result, axion creation proceeds in the regime, which is qualitatively similar to a narrow parametric resonance, cf. Refs.~\cite{Harigaya:2019qnl, Nakayama:2021avl}. Typically, efficiency of a narrow parametric resonance is limited by the Universe's expansion~\cite{Kofman:1997yn, Moroi:2020bkq} diluting enhancement due to Bose condensation of 
produced particles. 
Such a limitation does not take place in our case: in the radiation-dominated Universe, the time-decrease of Peccei-Quinn scale effectively compensates for the redshift of axion physical momenta. 
As a result, axion abundance grows exponentially and quickly reaches values necessary to explain dark matter (Section~\ref{sec:darkmatter}). Notably, this can be fulfilled for essentially any value of low energy Peccei-Quinn scale $f^{0}_{PQ}$.

Besides axions, fluctuations of the radial field $|S|$ serve as a promising source of non-trivial phenomenology. Contrary to conventional scenarios, where the field $|S|$ is superheavy, in our case typical values of the radial field 
are lying in the keV- and MeV-range, and thus particles $|S|$ can be abundantly produced in the stellar cores or in supernovae explosions. Furthermore, excitations of the field $|S|$ generated along with axions during 
the parametric resonance, may impact the Big Bang Nucleosynthesis (BBN) and the Cosmic Microwave Background (CMB); see Subsection~\ref{sec:perturbative}. This impact is unacceptably large, if the energy density of radial field particles generated in this way is comparable to that 
of axions at production, and if these radial fluctuations survive till BBN. However, for the values $f^{0}_{PQ} \simeq \mbox{(a few)} \times10^{8}~\mbox{GeV}$, one can efficiently thermalize radial fluctuations in the plasma and avoid the conflict with BBN and CMB data (Subsection~\ref{sec:thermal}). 
Yet for so low values of $f^{0}_{PQ}$, in the near future our scenario can be probed with helio-/haloscopes~\cite{IAXO:2019mpb, TASTE:2017pdv}, detailed study of stellar evolution, or a combination of both.

\section{The Model}
\label{sec:themodel}

Dynamics of the complex field $S$ carrying Peccei--Quinn charge is described by the Lagrangian: 
\begin{equation}
\label{base}
{\cal L}=\partial_{\mu} S^{\dagger} \partial_{\mu} S+g^2 S^{\dagger} S \phi^{\dagger} \phi  -\lambda_S \left(S^{\dagger}  S\right)^2 \; .
\end{equation}
We assume the following properties of the scalar field (or scalar multiplet) $\phi$: it is in thermal equilibrium with hot primordial plasma and has a non-zero expectation value at low temperatures:
\begin{equation}
\langle \phi^{\dagger} \phi \rangle =\frac{v^2_{\phi}}{2} \qquad (T\lesssim v_{\phi}) \; .
\end{equation}
It is tempting to identify $\phi$ with the SM Higgs field, but we will show later that this option is quite restrictive, and therefore we choose to keep the discussion generic. Note that we do not introduce a bare tachyonic mass for the field $S$ into the Lagrangian~\eqref{base}. Nevertheless, Peccei-Quinn symmetry is spontaneously broken, once we fix the sign of the coupling between the fields $\phi$ and $S$: 
\begin{equation}
g^2>0 \; .
\end{equation}
With this choice, the field $S$ acquires a non-zero expectation value at low temperatures given by 
\begin{equation}
\label{vevcold}
f^{0}_{PQ} =\frac{1}{\sqrt{2\beta}} \cdot \frac{v_{\phi}}{g} \qquad (T \lesssim v_{\phi}) \; .
\end{equation} 
Here we introduced the notation
\begin{equation}
\label{beta}
\beta \equiv \frac{\lambda_S}{g^4} \; .
\end{equation}
Convenience of the parameter $\beta$ is as follows: while the self-interaction constant $\lambda_S$ and the portal coupling $g^2$ can be very small in the phenomenologically most interesting part of parameter space, the ratio~\eqref{beta} is typically 
close to unity (cf. Refs.~\cite{Emond:2021vts, Babichev:2021uvl, Ramazanov:2021eya}). The upperscript `$0$' in Eq.~\eqref{vevcold} means that we are working in the low temperature limit. Being interested in the regime $f^{0}_{PQ} \gg v_{\phi}$, we should require that $\lambda_S \ll g^2$. Such small values are theoretically consistent: the lower bound on $\lambda_S$ following from the stability of two-field system reads $\lambda_S \lambda_{\phi} \geq g^4/4$, where $\lambda_{\phi}$ is the self-interaction constant of the field $\phi$, so that 
\begin{equation}
\label{stability}
\beta \geq \frac{1}{4\lambda_{\phi}} \; .
\end{equation}
 From perturbative unitarity we have $\lambda_{\phi} \lesssim 1$ and hence $\beta \gtrsim 1$.

Supernovae observations impose a lower bound on the Peccei--Quinn scale $f^{0}_{PQ}$, i.e., $f^{0}_{PQ} \gtrsim 10^{8}~\mbox{GeV}$. As it follows from Eq.~\eqref{vevcold}, for not very large expectation values $v_{\phi}$, we 
deal with a feebly coupled Peccei--Quinn field $S$:
\begin{equation}
\label{boundone}
g \simeq \frac{7 \cdot 10^{-6}}{\sqrt{\beta}}  \cdot \left(\frac{v_{\phi}}{1~\mbox{TeV}} \right) \cdot \left(\frac{10^{8}~\mbox{GeV}}{f^{0}_{PQ}} \right)\; .
\end{equation}
Furthermore, the scalar $S$ is rather light: 
\begin{equation}
\label{boundtwo}
M^{0}_{S} =\sqrt{2\lambda_S} f^{0}_{PQ} =\frac{v^2_{\phi}}{\sqrt{2\beta} f^{0}_{PQ}} \approx \frac{7~\mbox{MeV}}{\sqrt{\beta}}  \cdot \left(\frac{v_{\phi}}{1~\mbox{TeV}}  \right)^2 \cdot \left(\frac{10^8~\mbox{GeV}}{f^{0}_{PQ}} \right)\; .
\end{equation}
This fact clearly demonstrates a drastic difference between our setup and more conventional axion models, where the expectation value of the 
field $S$ is set as a bare parameter in the Lagrangian and the mass $M_S$ is assumed to be of the order of Peccei-Quinn symmetry breaking scale $f_{PQ}$.

Most interesting for our further discussion is the behaviour of the field $S$ at temperatures $T \gg v_{\phi}$. Notably, the field $S$ is in the spontaneously broken phase also at those times. 
Its expectation value is sourced by thermal fluctuations of the scalar (multiplet) $\phi$ described by the dispersion:
\begin{equation}
\label{var}
\langle \phi^{\dagger} \phi \rangle (t) =\frac{NT^2(t)}{24} \qquad (T \gtrsim v_{\phi})  \; ,
\end{equation}
where $N$ is the total number of relativistic degrees of freedom associated with $\phi$ at temperatures $T \gtrsim v_{\phi}$; note that $N=4$ in the case of Higgs doublet. Using Eqs.~\eqref{base},~\eqref{beta}, and~\eqref{var} we get the time-dependent expectation value $f_{PQ} (t)$:
\begin{equation}
\label{vevhot}
f_{PQ} (t)=\sqrt{\frac{N}{24\beta}} \cdot \frac{T(t)}{g} \qquad (T \gtrsim v_{\phi}) \; ,
\end{equation}
which redshifts $\propto 1/R$ with the scale factor $R (t)$. The temperature $T$ can reach large values in the primordial Universe; given also that the constant $g$ is tiny, we end up with a picture of 
Peccei--Quinn symmetry breaking scale gliding from sub-Planckian values to relatively small values as the Universe cools down.

\section{Parametric resonance}
\label{sec:resonance}

The best-known way of producing axions is through the misalignment mechanism, which leads to the correct dark matter abundance in the range $f^{0}_{PQ} \simeq 10^{11}-10^{12}~\mbox{GeV}$. For much smaller/larger values of $f^{0}_{PQ}$, coherent axion oscillations triggered by vacuum realignment give a negligible/too large contribution to dark matter~\cite{Preskill:1982cy, Abbott:1982af, Dine:1982ah}.  
 Axions are also emitted by global topological defects~\cite{Davis:1986xc, Kawasaki:2014sqa}, i.e., cosmic strings and domain walls. The latter are formed by the Kibble-Zurek mechanism, provided that the complex Peccei-Quinn field is set to zero in the post-inflationary Universe (commonly by the thermal mass), and the symmetry gets spontaneously broken later on. In our case, the light field $S$ minimally coupled to gravity gets offset from zero by superhorizon perturbations amplified during inflation. Furthermore, the thermal mass of the field $S$ acquired through the interaction with the scalar multiplet $\phi$ is tachyonic and thus cannot stabilise the field $S$ at zero. We conclude that global strings are not formed in our scenario\footnote{One can avoid this issue by assuming the coupling of the field $S$ to the Ricci scalar $\sim {\cal R} |S|^2$, but we choose to proceed with the minimal coupling 
in what follows.}. 

Time-dependence of the Peccei--Quinn scale enables another axion production mechanism through the decay of the field $|S|$ coherent oscillations in the parametric resonance regime. These oscillations 
occur because the field $S$ is initially displaced from its expectation value $f_{PQ}$ in the post-inflationary Universe. Axions produced in this decay can constitute all of the dark matter in the Universe for virtually any value of 
$f^{0}_{PQ}$, in particular as low as $f^{0}_{PQ} \lesssim 10^{11}~\mbox{GeV}$, for which the misalignment mechanism is inefficient. Notably, it is possible to describe the decay semi-analytically, at least in the small amplitude regime. 
We will demonstrate that the transfer of energy from the field $S$ oscillations to axions and potentially to radial fluctuations 
occurs very quickly within a few tens of Hubble times, and it is not affected by the Universe's expansion.

Let us first discuss the evolution of Peccei--Quinn field $S$ prior to the oscillatory phase. During inflation, the scalar $S$, which can be regarded as massless, acquires a nearly homogeneous value $\sqrt{\langle |S|^2 \rangle} \simeq 
\sqrt{{\cal N}_e} H_{inf}/(2\pi)$. Here $H_{inf}$ is the characteristic inflationary Hubble rate, and ${\cal N}_e$ is the total number of e-folds elapsed from the beginning of inflation. In what follows, we assume that this value of the field $S$ generated by the inflationary mechanism is well below the Planck mass $M_{Pl} \approx 2.44 \cdot 10^{18}~\mbox{GeV}$. After inflation, the field $S$ experiences a slow roll, during which its value does not change appreciably. This continues until the moment of time $t_i$ defined from $M_{S, i} \cdot t_i \simeq1$, or equivalently $M_{S, i} \simeq H_i$, when it starts oscillating. Taking into account that $M_{S, i} =\sqrt{2\lambda_S} f_{PQ, i}$ and using Eq.~\eqref{vevhot}, we can obtain the Universe's temperature at the time $t_i$:
\begin{equation}
\label{temponset}
T_i \simeq \sqrt{\frac{15N}{2\pi^2 g_* (T_i)}} \cdot g \cdot M_{Pl} \; .
\end{equation}
Substituting Eq.~\eqref{temponset} into Eq.~\eqref{vevhot}, we get
\begin{equation}
\label{nearPlanck}
f_{PQ, i} \simeq   \frac{\sqrt{5} N M_{Pl}}{4\pi \sqrt{\beta g_* (T_i)}}\; .
\end{equation} 
For $\beta$ not much greater than unity, which is most interesting phenomenologically (as we will see in what follows), the expectation value $f_{PQ, i}$ is only $1-2$ orders of magnitude below the Planck mass. We crucially 
assume in what follows that the value $f_{PQ, i}$ is larger than the field $|S|$ value at the onset of oscillations: 
\begin{equation}
\label{rel}
|S_i| \lesssim f_{PQ, i} \; .
\end{equation}
This warrants that the amplitude of oscillations is of the order of $f_{PQ}$ (baring fine-tuning between $|S_i|$ and $f_{PQ, i}$). Consequently, the 
initial energy density stored in the field $S$ oscillations is given by 
\begin{equation}
\label{energyin}
{\cal E}_i \simeq \frac{\lambda_S f^4_{PQ, i}}{4} \simeq \frac{M^2_{S, i} f^2_{PQ, i}}{8} \; .
\end{equation}
Note that $M_S \equiv \sqrt{2\lambda_S} f_{PQ}$ here is defined as the mass of the radial field $|S|$ fluctuations in the small amplitude case. 
We proceed with this definition of $M_S$ keeping in mind that the effective mass $M_{eff}$ of the field $|S|$ can be different from $M_S$ in the generic case.

 In the spontaneously broken phase, we decompose the complex field $S$ into a radius and a phase: 
\begin{equation}
S=\frac{1}{\sqrt{2}}\rho e^{i\frac{a}{f_{PQ}}} \; .
\end{equation}
Here $\rho =\sqrt{2} |S|$ and $a$ are the canonically normalized radial and axion fields, respectively. Before digging into details of axion and $\rho$-particle production, let us make one observation, which will greatly simplify our analysis. Namely, the action for the complex field $S$ with a time-decreasing expectation value in the radiation-dominated Universe can be written in the form: 
\begin{equation}
\label{actionflat}
{\cal S}=\int d\tau d{\bf x} \left[\frac{1}{2} \eta^{\mu \nu} \partial_{\mu} \tilde{\rho} \partial_{\nu} \tilde{\rho} +\frac{1}{2} \cdot \frac{\tilde{\rho}^2}{\tilde{f}^2_{PQ}}  \cdot \eta^{\mu \nu}\partial_{\mu} \tilde{a} \partial_{\nu} \tilde{a} -\frac{1}{4} \lambda_S\left( \tilde{\rho}^2-\tilde{f}^2_{PQ} \right)^2 \right] \; ,
\end{equation}
where $\eta_{\mu \nu}$ is the Minkowski metric related to Friedman-Lema\^itre-Robertson-Walker (FLRW) metric by $\eta_{\mu \nu} = g_{\mu \nu}/R^2(t)$. Recall that $R(t)$ denotes the Universe scale factor; the time derivative in Eq.~\eqref{actionflat} is with respect to the conformal time $d\tau=dt/R(t)$. We made the change of variables 
\begin{equation}
\tilde{a} =a \cdot R \qquad  \tilde{\rho}= \rho \cdot R \qquad \tilde{f}_{PQ} =f_{PQ} \cdot R\ . 
\end{equation}
We end up with an important conclusion that dynamics of the fields $\rho$ and $a$ during radiation-domination in the scenario with $f_{PQ} (t) \propto T (t)$ is equivalent to dynamics of the 
fields $\tilde{\rho}$ and $\tilde{a}$ in the Minkowski space-time with a constant $\tilde{f}_{PQ}$, cf. Ref.~\cite{Emond:2021vts}. Note that the stress energy tensors calculated with the flat spacetime metric $\eta_{\mu \nu}$
and FLRW metric are related by $\tilde{T}^{\mu \nu} =R^2 T^{\mu \nu}$. As $\tilde{T}^{0}_{0}$ is constant, the energy density stored in the field $S$ redshifts as
\begin{equation}
{\cal E} \equiv T^{0}_{0} \propto \frac{1}{R^4} \; ,
\end{equation} 
i.e., it behaves as radiation-like fluid.

Equations of motion for the fields $\tilde{\rho}$ and $\tilde{a}$ following from the action~\eqref{actionflat} are given by 
\begin{equation}
\label{radiusfull}
\partial_{\mu} \partial^{\mu} \tilde{\rho} +\frac{\tilde{\rho}}{\tilde{f}^2_{PQ}} (\partial_{\mu} \tilde{a})^2+ \lambda_S \cdot \left[ \tilde{\rho}^2 - \tilde{f}^2_{PQ}  \right] \cdot \tilde{\rho}=0 \; ,
\end{equation}
and 
\begin{equation}
\label{phasefull}
\partial_{\mu} \partial^{\mu}\tilde{a} +\frac{2\partial_{\mu}\tilde{\rho}}{\tilde{\rho}} \partial^{\mu} \tilde{a} = 0 \; ,
\end{equation}
respectively (raising and lowering of indices is accomplished with Minkowski metric). Note that the Hubble rate does not enter these equations in agreement with the above discussion: evolution of the fields $\tilde{\rho}$ and $\tilde{a}$ effectively proceeds in the flat spacetime. To make our further analysis tractable, we assume for a while that the field $\tilde{\rho}$ oscillates around the minimum with a small amplitude $C \ll \tilde{f}_{PQ}$. In that case, neglecting the backreaction of produced axions, we get from Eq.~\eqref{radiusfull}:
\begin{equation}
\tilde{\rho}=\tilde{f}_{PQ}+C \cdot \sin \left(\tilde{M}_{S} \tau \right) \; ,
\end{equation}
where $\tilde{M}_{S}=M_{S} R =\sqrt{2\lambda_S} \tilde{f}_{PQ}$ is the constant rescaled mass of the complex field $S$. With no loss of generality, the moment $\tau=0$ corresponds to the first minimum crossing of the field $\rho$.
A more realistic scenario with the large initial amplitude of oscillations $C \sim \tilde{f}_{PQ}$ leads to qualitatively same results, and we discuss it later on in this section.

We write the axion field as a combination of creation and annihilation 
operators $A^{\dagger}_{{\bf k}}$ and $A_{{\bf k}}$: 
\begin{equation}
\delta \tilde{a} (\tau, {\bf x})= \int \frac{d{\bf k}}{(2\pi)^{3/2}} \left[e^{i{\bf kx}} \delta \tilde{a} (\tau, {\bf k}) A_{{\bf k}} +e^{-i {\bf kx}} \delta \tilde{a}^{*}(\tau, {\bf k}) A^{\dagger}_{{\bf k}}  \right] \; ,
\end{equation}
where $\delta \tilde{a} (\tau, {\bf k})$ is a mode function, and $\delta \tilde{a}^{\dagger} (\tau, {\bf k})$ is its complex conjugate. The analogous decomposition into a set of (different) creation and annihilation operators can be performed 
for the radial field perturbations with an obvious replacement of the mode function $\delta \tilde{a} (\tau, {\bf k}) \rightarrow \delta \tilde{\rho} (\tau, {\bf k})$. 
For simplicity, we assume vacuum initial conditions corresponding to the following choice of mode functions: 
\begin{equation}
\label{vacuum}
\delta \tilde{a}(\tau, {\bf k})=\frac{1}{\sqrt{2k}} e^{-ik\tau} \qquad \delta \tilde{\rho} (\tau, {\bf k}) =\frac{1}{\sqrt{2\omega_k}} e^{-i\omega_{{\bf k}}\tau} \; ,
\end{equation}
where $\omega_k =\sqrt{k^2+\tilde{M}^2_{S}}$. Note that generically the modes of interest exit the horizon at the last stages of inflation, which leads to their efficient amplification compared to the expressions~\eqref{vacuum}. This ambiguity about initial conditions is irrelevant: as soon as the mode functions $\delta \tilde{a}(\tau, {\bf k})$ grow exponentially fast in the parametric resonance regime, information about initial 
conditions is erased within a few oscillation periods. Shortly we will see that this is indeed the case.

As it follows from Eq.~\eqref{radiusfull}, axionic mode functions $\delta \tilde{a} (\tau, {\bf k})$ comply with the equation:
\begin{equation}
\label{phasefullmode}
\delta \tilde{a}''  +k^2 \delta \tilde{a}  +\frac{2\tilde{\rho}'}{\tilde{\rho}} \delta \tilde{a} ' =0 \; .
\end{equation}
In the small amplitude regime, it can be rewritten as
\begin{equation}
\label{sa}
\delta \tilde{a}''+k^2 \delta \tilde{a} +\frac{2C \tilde{M}_{S}}{\tilde{f}_{PQ}} \cos \left(\tilde{M}_{S} \tau \right) \delta \tilde{a}'=0 \; . 
\end{equation}
We observe that upon a change of variables: 
\begin{equation}
\delta \tilde{a} (\tau, {\bf k}) ={\cal Y} (\tau, {\bf k}) \cdot \mbox{exp} \left[-\int^{\tau}_0 d\tau' \frac{\tilde{M}_{S} C}{\tilde{f}_{PQ}} \cos \left(\tilde{M}_{S} \tau' \right) \right] \; ,
\end{equation}
Eq.~\eqref{sa} takes the form of Mathieu equation:
\begin{equation}
\label{Mathieu}
\frac{d^2 {\cal Y}}{dz^2_a}+ \left[Q_a+2q_a \sin (2z_a) \right] {\cal Y} =0 \; ,
\end{equation}
where 
\begin{equation}
z_a =\frac{\tilde{M}_{S} \tau}{2} \quad Q_a=\frac{4k^2}{\tilde{M}^2_{S}} \quad q_a=\frac{2C}{\tilde{f}_{PQ}} \; .
\end{equation}
Our assumption $C \ll \tilde{f}_{PQ}$ implies that $q_a \ll 1$, which corresponds to the narrow parametric resonance. In this regime, only modes of the field ${\cal Y}$ and hence $\delta \tilde{a}$ with conformal momenta $k \approx \tilde{M}_S/2$ experience an instability leading to their exponential growth:
\begin{equation}
\label{narrowgrowth}
\delta \tilde{a} \propto \mbox{exp} \left(\frac{q_a \tilde{M}_{S} \tau}{4} \right) \; .
\end{equation}
On the other hand, away from the main resonance band centered at $k=\tilde{M}_S/2$, the growth is very modest, and we can safely neglect these modes. Using the estimate $M_{S, i} \simeq H_i$, we conclude that physical wavelengths of produced massless particles $a$ are comparable with the horizon size, at least at the initial stage of parametric resonance decay. In this regard, the latter is similar to the axion production mechanism through the decay of global cosmic strings. 

\begin{figure}[tb!]
  \begin{center}
    \includegraphics[width=\columnwidth,angle=0]{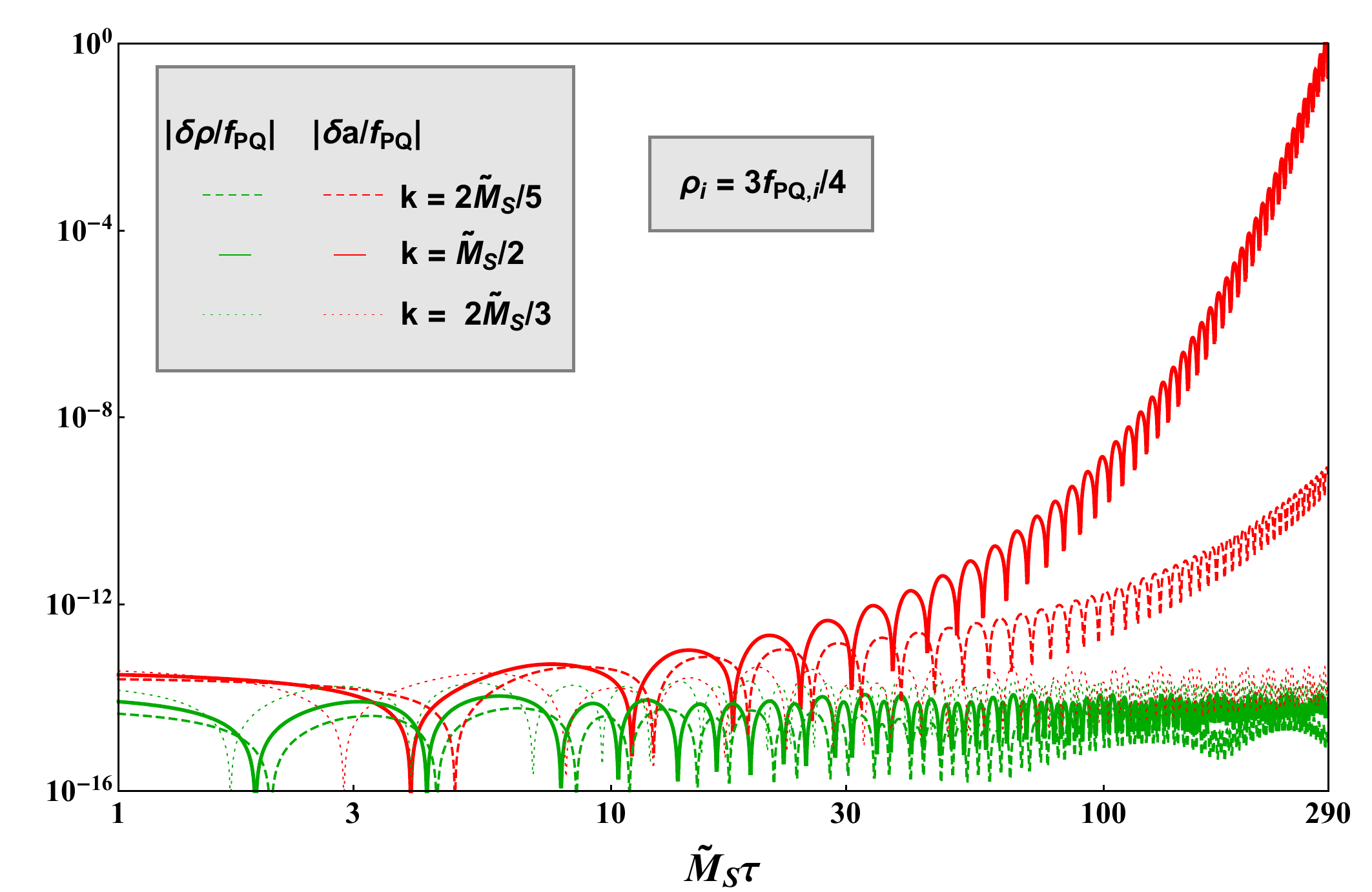}
  \caption{Evolution of perturbations $\delta a=\delta \tilde{a}/R$ and $\delta \rho=\delta \tilde{\rho}/R$ is shown in absolute values for different conformal momenta $k$ in the narrow parametric resonance regime neglecting backreaction of produced particles on the oscillating field. The perturbations $\delta a$ and $\delta \rho$ 
  are normalized to the expectation value $f_{PQ}$ of the Peccei--Quinn field $S$. Vacuum initial conditions~\eqref{vacuum} have been assumed. We have set the model constants to values $g=10^{-7}$ and $\beta =10$. The simulation is stopped 
  at the point, when the whole initial energy of the coherently oscillating radial field is transferred to axion fluctuations, corresponding to $|\delta a | \sim f_{PQ}$.}\label{narrowresonance}
  \end{center}
\end{figure}

The evolution of axion modes $\delta a$ in the narrow parametric resonance is shown in Fig.~\ref{narrowresonance}. Indeed, they grow extremely fast (for certain conformal momenta) 
from minuscule vacuum values to huge values of the order $\delta a \simeq f_{PQ}$, -- in accordance with Eq.~\eqref{narrowgrowth}. We would like to stress that the narrow parametric resonance is extremely efficient in our case for two major reasons. First, the system effectively evolves in the Minkowski spacetime; as a result, the narrow resonance band at $k \approx \tilde{M}_S/2$ remains stable with time. To paraphrase, the mass $M_S$ redshifts in the same way as physical momenta. On the other hand, in the case of constant $f_{PQ}$, the mass $M_S$ remains constant, while physical momenta redshift. Consequently, the resonance band gets diluted and enhancement by Bose condensate of produced particles is considerably reduced. This clearly demonstrates the advantage of having a temperature-dependent Peccei-Quinn scale. Second, axions have negligible interactions with the 
thermal plasma. This again prevents their escape from the resonance band. 

In Fig.~\ref{narrowresonance}, we also show evolution of radial modes, which are generically excited by oscillations of the radius $\tilde{\rho}$ through the quartic self-interaction. The equation of motion describing evolution of mode functions $\delta \tilde{\rho} (\tau, {\bf k})$ is given by
\begin{equation}
\label{genradpert}
\delta \tilde{\rho}''+k^2 \delta \tilde{\rho}+\lambda_S \cdot \left[3\tilde{\rho}^2-\tilde{f}^2_{PQ} \right] \delta \tilde{\rho}=0 \; .
\end{equation}
In the small amplitude regime, the latter again reduces to Mathieu equation~\eqref{Mathieu}, where one should make a replacement ${\cal Y} \rightarrow \delta \tilde{\rho}$, change subscripts $a \rightarrow \rho$, and identify 
\begin{equation}
z_{\rho}=\frac{\tilde{M}_{S} \tau}{2}+\frac{\pi}{2} \qquad Q_{\rho} =4 \left(\frac{k^2}{\tilde{M}^2_{S}}+1\right) \qquad q_{\rho} =\frac{6C}{\tilde{f}_{PQ}} \; .
\end{equation}
As it follows from Fig.~\ref{narrowresonance} production of radial perturbations is much less efficient compared to the case of axions. This can be simply explained by the fact that the narrow parametric resonance is prominent only in the situation, when the perturbative decay is allowed~\cite{Gorbunov}. Clearly, the decay of the coherent field into radial excitations triggered by the quartic self-interaction is perturbatively prohibited.  

\begin{figure}[tb!]
  \begin{center}
    \includegraphics[width=0.48\columnwidth,angle=0]{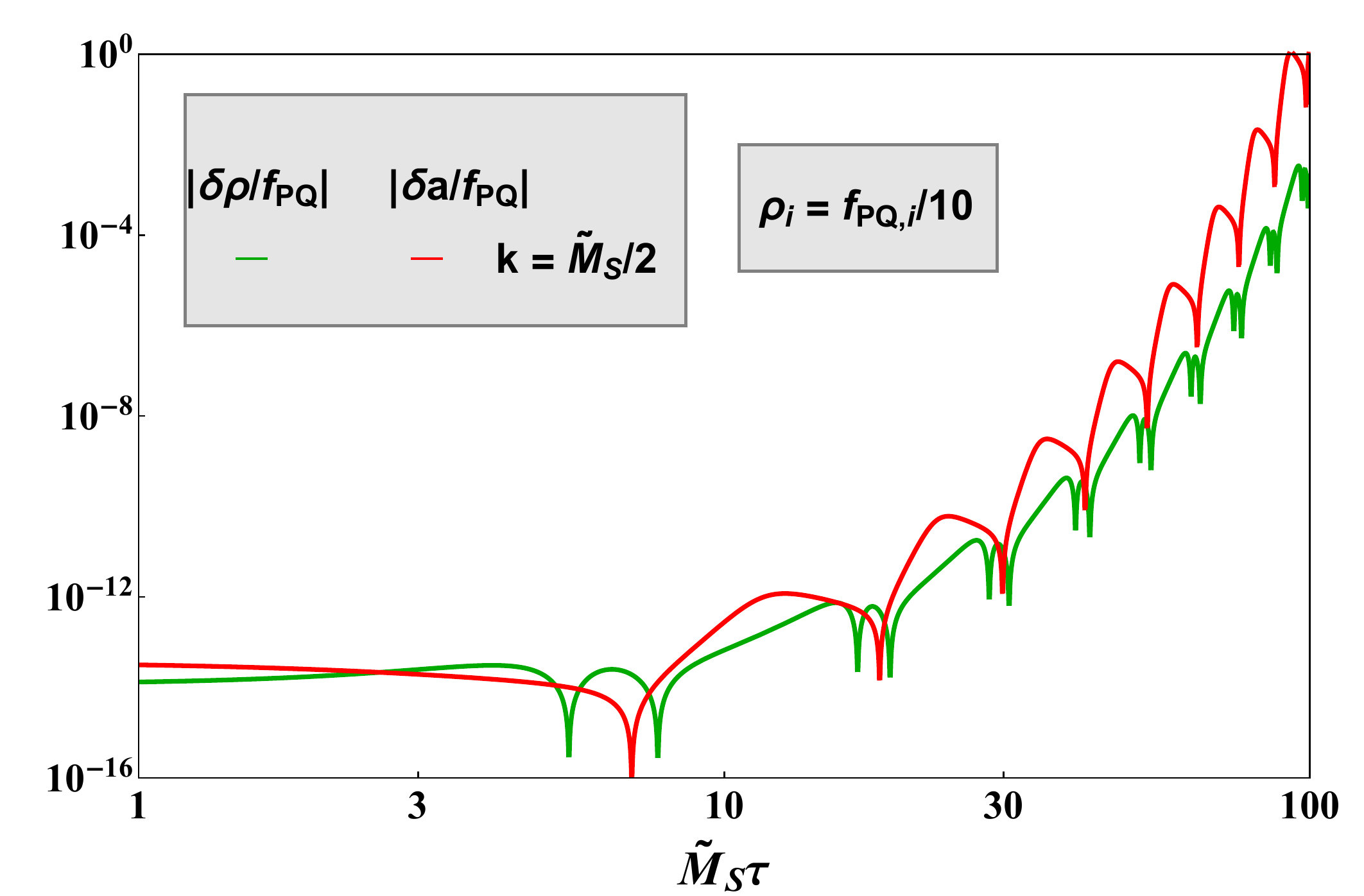}
     \includegraphics[width=0.48\columnwidth,angle=0]{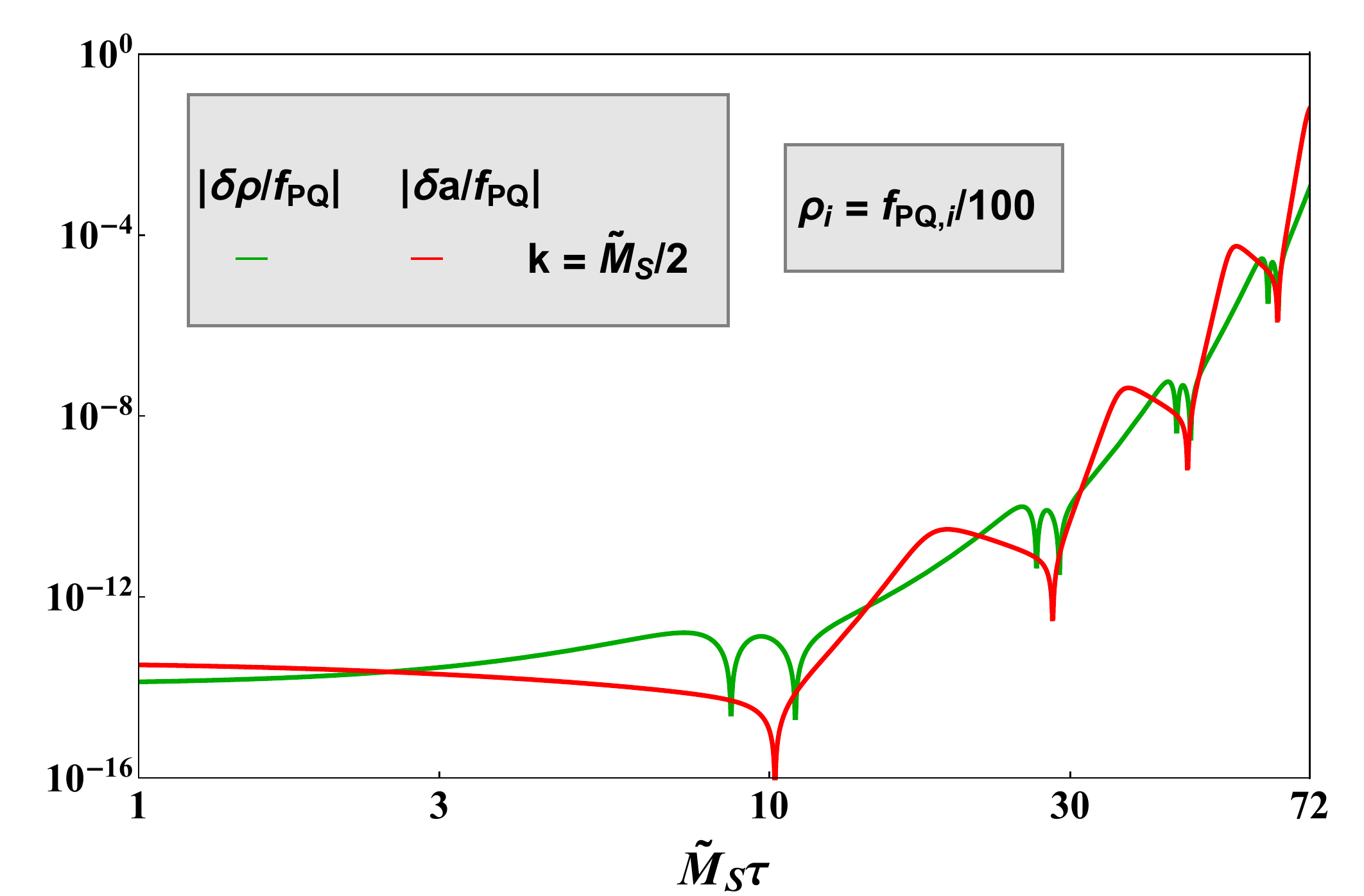}
      \includegraphics[width=0.48\columnwidth,angle=0]{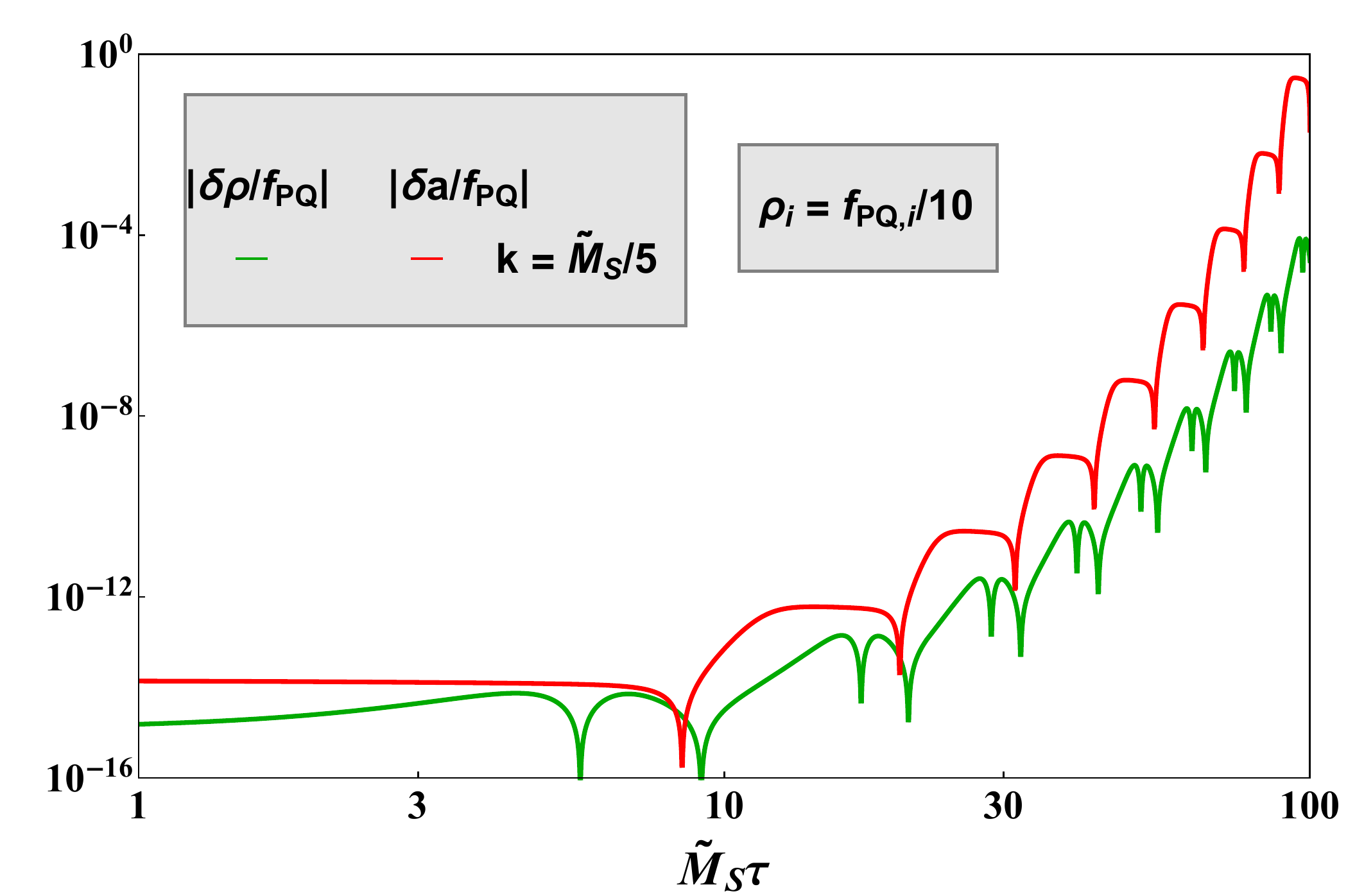}
       \includegraphics[width=0.48\columnwidth,angle=0]{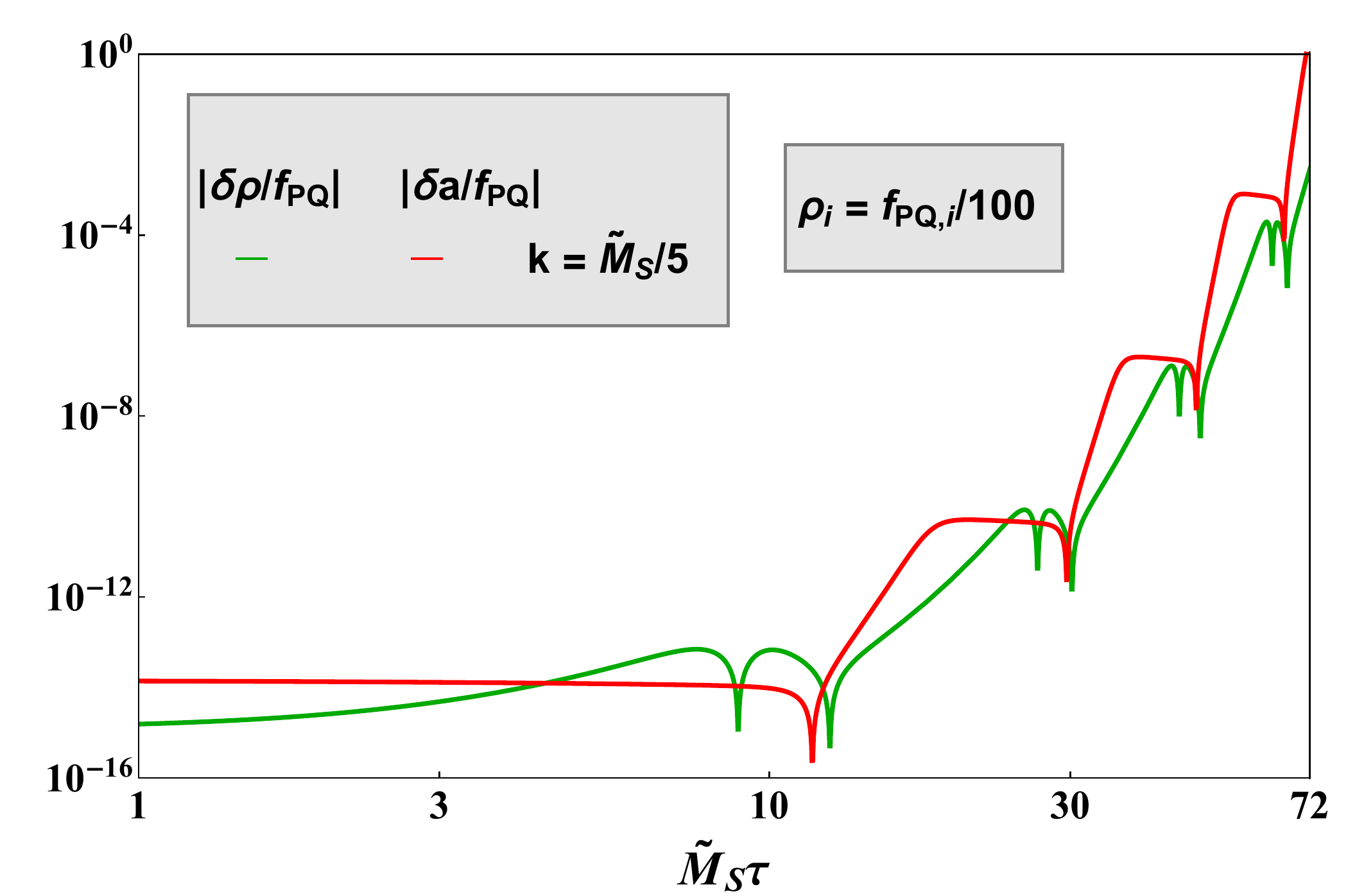}
       \includegraphics[width=0.48\columnwidth,angle=0]{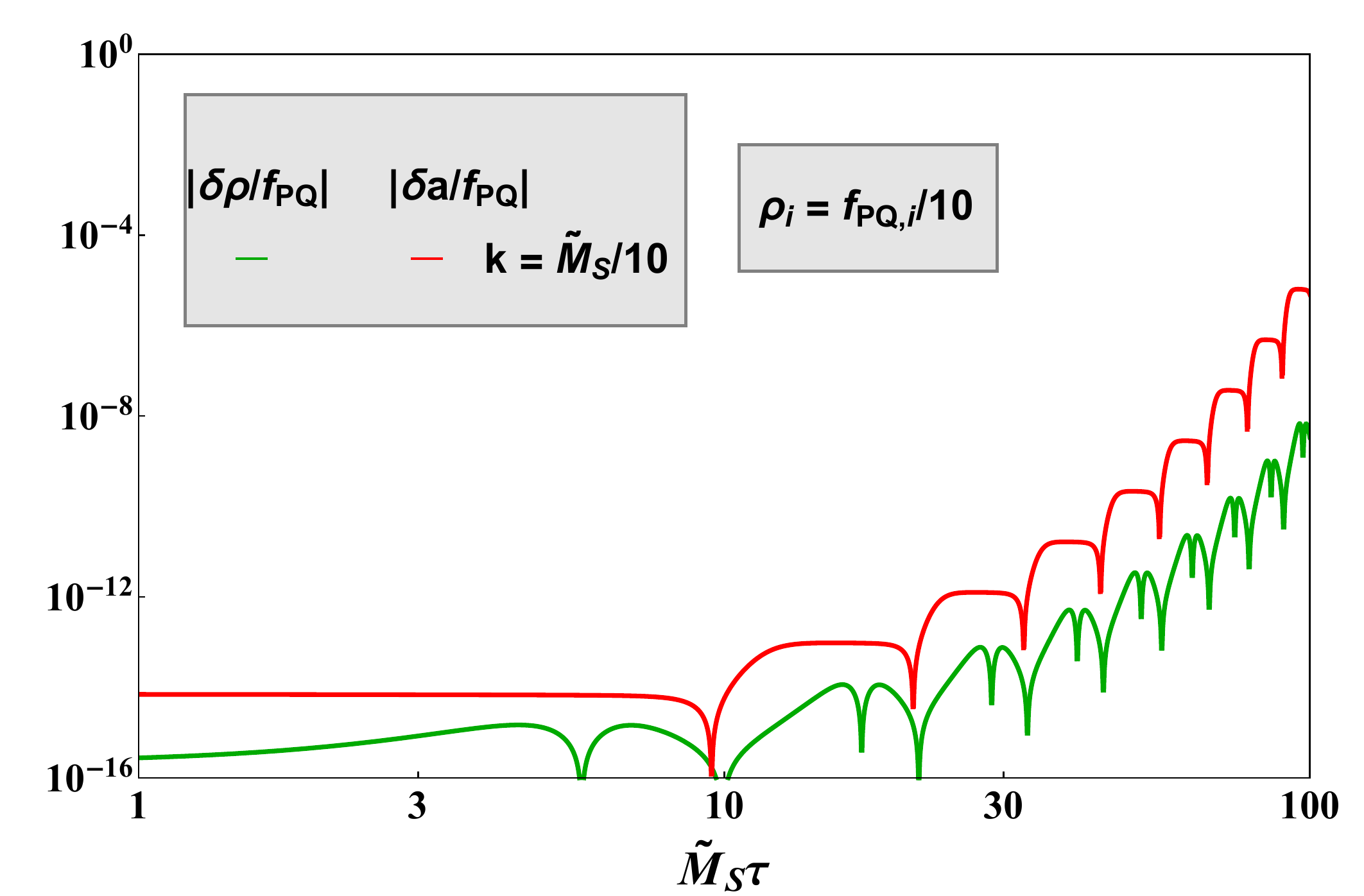}
        \includegraphics[width=0.48\columnwidth,angle=0]{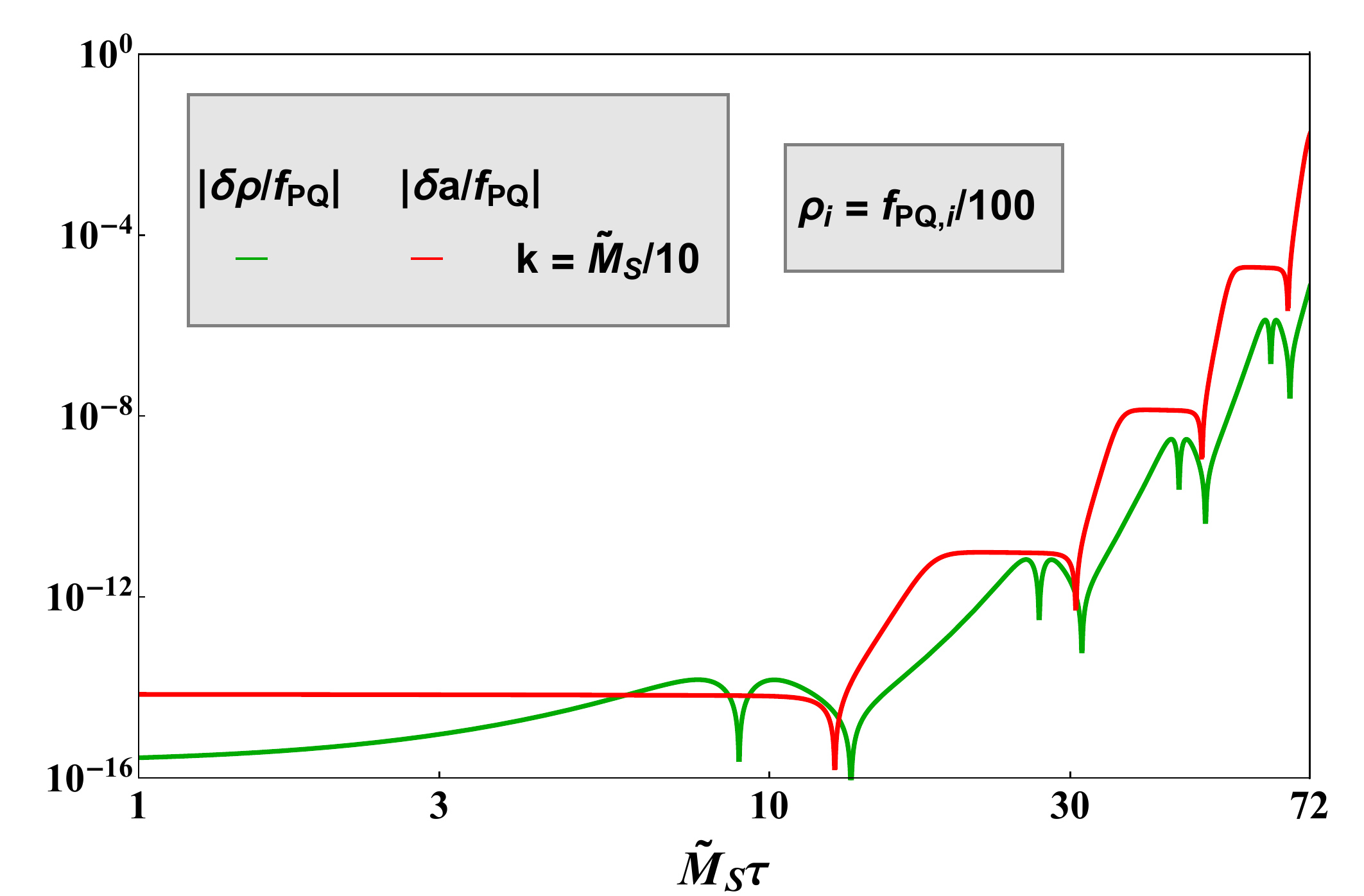}
  \caption{The same as in Fig.~\ref{narrowresonance}, but for small initial values of the radial field $\rho_i$, i.e., $\rho_i \ll f_{PQ, i}$.}\label{resonance}
  \end{center}
\end{figure}

{\it Natural case $C_i \sim f_{PQ, i}$}. The picture of narrow parametric resonance discussed above assumed a fine-tuning $\rho_i \sim f_{PQ, i}$. Recall that generically we expect the hierarchy $\rho_i \ll f_{PQ, i}$. Consequently, the initial amplitude of oscillations 
$C_i \simeq |f_{PQ, i} -\rho_i| \simeq f_{PQ, i}$ is quite large. This means that the above analysis must be corrected, not only because the condition $q \ll 1$ is not fulfilled anymore, but also since an approximation of the effective quadratic potential is invalidated, and oscillations 
become anharmonic. We numerically solve Eqs.~\eqref{radiusfull},~\eqref{phasefull}, and~\eqref{genradpert} for different values of the ratio $k/\tilde{M}_{S}$ and initial $\rho_i$, which is now chosen to be small relative to $f_{PQ, i}$. We again neglect backreaction by omitting the term $\propto (\partial_{\mu} \tilde{a})^2$ in Eq.~~\eqref{radiusfull}. The results are shown in Fig.~\ref{resonance}. As it follows, axion production in this case is qualitatively similar to what we have seen in the narrow parametric resonance regime, but it occurs in a broader band and it is more prominent for modes with slightly lower momenta $k <\tilde{M}_S/2$.

At the same time, production of radial modes also shown in Fig.~\ref{resonance}, is much more efficient compared to the small amplitude case and depends on the initial value of the radius $\rho_i$. Note that the above argument based on the possibility of the perturbative decay is not applicable anymore, as we do not deal with the narrow parametric resonance. We observe that for $\rho_i/f_{PQ}  \rightarrow 0$, production of radial modes may become non-negligible or even comparable to that of axions (modulo possible backreaction effects discussed below), and thus cannot be ignored. 
Therefore, we introduce the parameter
\begin{equation}
\label{ratio}
r \equiv \frac{{\cal E}_{\rho}}{{\cal E}_{\rho}+{\cal E}_{a}} \; ,
\end{equation}
where the energy density ${\cal E}_a$ (${\cal E}_{\rho}$) of the particles $a$ ($\rho$) is defined by the end of parametric resonance. We would like to stress that $r$ is not a model parameter, but simply takes into account our ignorance 
about the initial amplitude of oscillations and backreaction issues. Depending on the actual value of the parameter $r$, particles $\rho$ may strongly impact dark matter implications of the model as well as predictions for the  BBN and the CMB. In what follows, 
we will assume that the parameter $r$ does not exceed unity significantly, i.e., $r \lesssim 1$, but can take an arbitrary value otherwise.

{\it Backreaction.} While a careful analysis of backreaction is out of the scope of the present work, below we make some speculations about the final stages of parametric resonance. 
When backreaction due to particles $a$ and $\rho$ becomes large enough, the amplitude of the field $\rho$ oscillations drops significantly. The time-dependence of the amplitude $C$ is expected to strongly affect further particle production. 
In particular, less the amplitude $C(t)$ is, better is the approximation of the narrow parametric resonance. As we have discussed above, production of particles $\rho$ is essentially shut off for small $C$, while production 
of axions is still efficient. This suggests that the number density of axions significantly exceeds that of particles $\rho$ by the end of parametric resonance.

On the other hand, particles $\rho$ may appear due to axions rescattering off the condensate $\rho(t)$. Let us argue, however, that rescattering can be inefficient in our case. For this purpose, we first estimate the maximal amplitude $\delta \tilde{a}_{max}$ of axionic fluctuations formally assuming that the whole energy of oscillations is eventually transferred to the energy of freely propagating axions. From the energy conservation, we have
\begin{equation}
\label{maximal}
 \frac{\tilde{M}^2_{S} (\delta \tilde{a} )^2_{max}}{4} \simeq \frac{\lambda_S \tilde{f}^4_{PQ}}{4} \simeq \frac{\tilde{M}^2_{S} \tilde{f}^2_{PQ}}{8} \; .
\end{equation}
Here we took into account that most of energy is contained in the mode with $k \approx \tilde{M}_{S}/2$. Hence, the energy density of axions is estimated as 
\begin{equation}
{\cal E}_a =\frac{(\partial_{0} \tilde{a})^2+(\partial_i \tilde{a})^2}{2} \simeq \frac{\tilde{M}^2_{S} (\delta \tilde{a})^2}{4} \; .
\end{equation}
From Eq.~\eqref{maximal}, we obtain
\begin{equation}
\delta \tilde{a}_{max} \simeq \frac{\tilde{f}_{PQ}}{\sqrt{2}} \; .
\end{equation}
 The maximal correction to Eq.~\eqref{radiusfull} caused by produced axions is estimated as 
\begin{equation}
\label{correction}
\frac{\langle (\partial_{\mu} \tilde{a} )^2 \rangle}{\tilde{f}_{PQ}} \simeq \frac{q \tilde{M}^2_{S} \tilde{f}_{PQ}}{8} \; .
\end{equation}
Note that one has $\langle (\partial_{\mu} \tilde{a})^2 \rangle =0$ in the limit $q \rightarrow 0$. This explains appearance of the factor $q$ on the r.h.s. of Eq.~\eqref{correction}. 
We observe that the correction~\eqref{correction} does not exceed other terms in Eq.~\eqref{radiusfull}, i.e., 
\begin{equation}
\tilde{M}^2_{S} \delta \tilde{\rho} \simeq \frac{q\tilde{ M}^2_{S} \tilde{f}_{PQ}}{2} \; .
\end{equation}
This means that the effect of the axion field on the frequency of radial field oscillations can be negligible well until the time, when the axion field fluctuation reaches the maximum value, cf. Ref.~\cite{Kofman:1997yn}. 
Hence, there are good reasons to expect that most of the energy density of the radial field oscillations transfers into axions, i.e., $r \ll 1$. On the other hand, the situation is borderline, and a rigorous analysis is required.

In the end of this section, we would like to stress that axion production through the parametric resonance has been previously studied in the literature. The crucial distinction of our analysis from the early works~\cite{Co:2017mop, Co:2020dya, Harigaya:2019qnl, Nakayama:2021avl, Ema:2017krp}, is the temperature-dependence of the Peccei-Quinn scale $f_{PQ} (t) \propto T(t)$. This affects not only parametric resonance, but also later evolution of axions and $\rho$-particles, as it is discussed in the following sections. 
Furthermore, Refs.~\cite{Co:2017mop, Co:2020dya} assume a large 
initial amplitude of the field $\rho$ oscillations $\rho_i \gg f_{PQ}$. i.e., the inequality~\eqref{rel} is grossly violated. Such a large $\rho_i$ is achieved due to the Hubble-induced mass during inflation common in supersymmetric models. 
As a result, particle production effectively proceeds in the quartic potential, at least at early stages, cf. Ref.~\cite{Greene:1997fu}. On the other hand, in our case the potential is better approximated as a quadratic one. 
Note also that Refs.~\cite{Co:2017mop, Co:2020dya} assume comparable number densities of axions and $\rho$-particles produced by the end of parametric resonance. Instead, we do not fix the ratio~\eqref{ratio} allowing it to be very small, in accordance with the discussion above.

\section{Axions as dark matter}
\label{sec:darkmatter}

In this section, we define the region of parameter space, where the particles $a$ produced during the resonance decay of the 
radial field $\rho$, constitute the main component of cold dark matter in the Universe. Initially massless, the axion $a$ acquires a small mass $m_a$ due to QCD instanton effects. At temperatures below the 
QCD phase transition, the mass $m_a$ reads
\begin{equation}
m_{a} \approx 5 \cdot 10^{-2}~\mbox{eV} \cdot \left( \frac{10^8~\mbox{GeV}}{f^{0}_{PQ}} \right) \; ,
\end{equation}
or equivalently 
\begin{equation}
\label{axionmass}
m_a \approx \frac{g \cdot \beta \cdot (100~\mbox{MeV})^2}{\sqrt{2} v_{\phi}} \; .
\end{equation}
Recall that $f^{0}_{PQ}$ is the low energy Peccei--Quinn scale given by Eq.~\eqref{vevcold}. 
Let us make an important remark here. Writing Eq.~\eqref{axionmass}, we implicitly ignore the difference between the scale of Peccei--Quinn symmetry breaking $f_{PQ}$ and the axion decay constant $f_a$. 
Generically, however, this is not the case: the scales $f_a$ and $f_{PQ}$ can differ by a factor of few, i.e., $f_a=f_{PQ}/(2{\cal N})$, where ${\cal N}$ is the QCD anomaly factor. 
While in the simplest KSVZ scenario ${\cal N}=1/2$, and one indeed has $f_{PQ}=f_a$, DFSZ models predict ${\cal N}=3$, so that $f_a=f_{PQ}/6$. Still, 
unless the opposite is stated, we stick to the choice $f_a =f_{PQ}$ in what follows, which is also the best-motivated option from the viewpoint of axion domain wall problem~\cite{Sikivie:1982qv}. 

Axions become non-relativistic at the time $t_{\times}$ defined from 
\begin{equation}
\label{turn}
\frac{k}{R_{\times}} \simeq m_{a} \; ,
\end{equation}
and since then contribute to dark matter. Fitting the observed relic abundance of dark matter with sufficiently cold axions, imposes important constraints on our model parameter space. 
Let us first obtain the Universe's temperature $T_{\times}$ at the time $t_{\times}$. During the parametric resonance decay of the Peccei-Quinn field $S$, axions are produced with characteristic momenta $k/R \simeq M_{S}/2$, so that
\begin{equation}
\label{long}
\frac{k}{R_{\times}} = \frac{k}{R_i} \cdot \frac{R_i}{R_{\times}} \simeq \frac{M_{S} (t_i)}{2} \cdot \frac{T_{\times}}{T_i} \cdot \left(\frac{g_* (T_{\times})}{g_* (T_i)} \right)^{1/3}  \; .
\end{equation}
Here we took into account entropy conservation in the comoving volume. Substituting $M_S (t_i) =\sqrt{2\lambda_S} f_{PQ, i}$, using Eqs.~\eqref{beta} and~\eqref{vevhot}, we get
\begin{equation}
\label{newlong}
\frac{k}{R_{\times}} \simeq 0.15 \sqrt{N} \cdot g T_{\times}  \cdot \left(\frac{g_* (T_{\times})}{g_* (T_i)}  \right)^{1/3}  \; .
\end{equation}
Combining Eqs.~\eqref{axionmass},~\eqref{turn} and~\eqref{newlong}, we obtain
\begin{equation}
\label{Tturn}
T_{\times} \simeq 50~\mbox{keV} \cdot \sqrt{\frac{\beta}{N}} \cdot \left(\frac{g_* (T_i)}{g_* (T_{\times})} \right)^{1/3} \cdot \frac{1~\mbox{TeV}}{v_{\phi}} \; . 
\end{equation}
In what follows, we assume that most of the energy density of the oscillating Peccei-Quinn complex field is transferred to axions, i.e., $r\lesssim 1/2$ in Eq.~\eqref{ratio}. Consequently, the energy density of axion dark matter is given by 
\begin{equation}
\label{axenergy}
{\cal E}_{a} (t) \simeq \frac{\lambda_S v^4_S (t_i)}{4} \cdot \left(\frac{R_i}{R_{\times}} \right)^4 \cdot \left(\frac{R_{\times}}{R (t)} \right)^3 \simeq \frac{4.5 \cdot 10^{-4} N^2 g^{1/3}_{*} (T_{\times}) g_* (T) }{g^{4/3}_{*} (T_i)  \beta} \cdot 
T_{\times} T^3 (t) \; ,
\end{equation}
where we used Eqs.~\eqref{vevhot} and~\eqref{energyin}. The requirement that axions make all of dark matter in the Universe imposes the following condition at the time $t=t_{eq}$ of matter-radiation equality: 
\begin{equation}
\label{eq}
{\cal E}_{a} (t_{eq}) \simeq {\cal E}_{rad} (t_{eq}) \simeq \frac{\pi^2 g_* (T_{eq}) T^4_{eq}}{30} \; ,
\end{equation}
where ${\cal E}_{rad} (t_{eq})$ and $T_{eq} \approx 0.8~\mbox{eV}$ are the radiation energy density and the Universe's temperature at the time $t_{eq}$, respectively. Substituting Eq.~\eqref{axenergy} into the condition~\eqref{eq} and using Eq.~\eqref{Tturn}, we obtain the value of the parameter $\beta$: 
\begin{equation}
\label{betaaxionDM}
\beta \simeq 40 \cdot   \left(\frac{N}{4} \right)^3 \cdot \left(\frac{100}{g_* (T_i)} \right)^2 \cdot \left(\frac{1~\mbox{TeV}}{v_{\phi}} \right)^2 \; .
\end{equation}
The latter estimate together with the condition~\eqref{stability} following from the stability in the system of two fields $S$ and $\phi$, leads to the important upper bound on the expectation value $v_{\phi}$: 
\begin{equation}
\label{uppervev}
v_{\phi} \lesssim \mbox{13~TeV} \cdot \sqrt{\lambda_{\phi}} \cdot \left(\frac{N}{4} \right)^{3/2} \cdot \left(\frac{100}{g_* (T_i)} \right) \; .
\end{equation}
Substituting Eq.~\eqref{betaaxionDM} into Eq.~\eqref{Tturn}, we fix the Universe's temperature $T_{\times}$, at which axions become non-relativistic:  
\begin{equation}
T_{\times} \simeq 0.4~\mbox{MeV} \cdot \left(\frac{N}{4} \right) \cdot \left(\frac{100}{g_* (T_i)} \right)^{2/3} \cdot \left(\frac{10}{g_* (T_{\times})} \right)^{1/3} \cdot \left(\frac{1~\mbox{TeV}}{v_{\phi}} \right)^2 \; .
\end{equation}
The temperature $T_{\times}$ should be limited from below: otherwise, axions may turn out to be too warm, and one risks to distort the bottom up picture of 
large scale structure formation. Conservatively, we enforce the constraint $T_{\times} \gtrsim 10~\mbox{keV}$, which can be converted into the upper bound on $v_{\phi}$:  
\begin{equation}
\label{uppervevtwo}
v_{\phi} \lesssim 6~\mbox{TeV} \cdot \sqrt{\frac{N}{4}} \cdot \left(\frac{100}{g_* (T_i)} \right)^{1/3} \cdot \left(\frac{10}{g_* (T_{\times})} \right)^{1/6} \; .
\end{equation}
Notably, this is very similar to the limit~\eqref{uppervev}, which follows from stability considerations. The bound in Eq.~\eqref{uppervevtwo} can be made more precise through the study of Lyman $\alpha$ absorption lines, but we do not expect a large difference compared to Eq.~\eqref{uppervevtwo}.

Finally, using Eqs.~\eqref{boundone} and~\eqref{boundtwo}, we can define the constant $g$ and the low temperature mass $M^{0}_S$ of the Peccei-Quinn field in our axion dark matter scenario. The former is given by
\begin{equation}
\label{g}
g \simeq  1.3 \cdot 10^{-6} \cdot \left(\frac{4}{N} \right)^{3/2} \cdot \left(\frac{g_* (T_i)}{100} \right)  \cdot \left(\frac{v_{\phi}}{1~\mbox{TeV}} \right)^2 \cdot \left(\frac{10^8~\mbox{GeV}}{f^{0}_{PQ}} \right) \; .
\end{equation}
This reiterates our point that we deal with a very weakly coupled field $S$. Substituting Eq.~\eqref{g} into $M^0_S=gv_{\phi}$, we also confirm that it is rather light: 
\begin{equation}
\label{masslow}
M^{0}_{S} \simeq 1.3~\mbox{MeV}  \cdot \left(\frac{4}{N} \right)^{3/2} \cdot \left(\frac{g_* (T_i)}{100} \right)  \cdot \left(\frac{v_{\phi}}{1~\mbox{TeV}} \right)^3 \cdot \left(\frac{10^8~\mbox{GeV}}{f^{0}_{PQ}} \right)\; .
\end{equation}
So small values of $M_S$ should be taken with caution, because generically they allow for production of particles $S$ in stellar cores~\cite{Hardy:2016kme,Knapen:2017xzo} (cf. Refs.~\cite{Ellis:1987pk, Co:2017orl}). 
This indeed happens, if the scalar $\phi$ is identified with the SM Higgs field $H$, i.e., $\phi=H$. Due to the mixing between the Higgs field with the Peccei - Quinn field $S$, the latter effectively couples to electrons and nucleons. 
The impact of these couplings on stellar evolution is commonly expressed in terms of the mixing angle between the fields $S$ and $H$ defined as 
\begin{equation}
\sin \theta =\frac{g^2 f^{0}_{PQ} v_{H}}{m^2_H} \; .
\end{equation}
Using Eqs.~\eqref{vevcold},~\eqref{g}, substituting Eq.~\eqref{betaaxionDM} and the values $v_H \approx 246~\mbox{GeV}$,~$m_H \approx 125~\mbox{GeV}$, we obtain 
\begin{equation}
\sin \theta \simeq 10^{-8} \cdot \left(\frac{10^{8}~\mbox{GeV}}{f^{0}_{PQ}} \right) \; .
\end{equation}
Comparing this with the limits shown in Fig.~6 of Ref.~\cite{Hardy:2016kme}, we observe that the values $f^{0}_{PQ} \sim 10^{8}~\mbox{GeV}$ (corresponding to masses $M_S \sim 20~\mbox{keV}$) and $f^{0}_{PQ} \gtrsim 3 \cdot 10^{9}~\mbox{GeV}$ 
(for which $\sin \theta \lesssim 3 \cdot 10^{-10}$) are allowed. The region $10^{8}~\mbox{GeV} \lesssim f^{0}_{PQ} \lesssim 3 \cdot 10^{9}~\mbox{GeV}$ is in tension with evolution of red giants caused by the coupling to electrons. 
Note that considerations of the next section based on consistency with cosmology favour values of $f^{0}_{PQ}$ below $10^{9}~\mbox{GeV}$. This paves the way for ruling out our scenario 
in the case of the SM Higgs field.

\section{Cosmology of radial excitations}
\label{sec:cosmology}

The detailed study of cosmological evolution of particles $\rho$ produced during parametric resonance is model-dependent; in particular, it requires knowledge of the field $\phi$-properties. 
For simplicity, we assume that the field $\phi$ does not have decay channels to other particle species.  In this case, we show that even a tiny fraction of particles $\rho$ produced during parametric resonance may have a sizeable impact on cosmological observations. This serves 
as a source of constraints on the model parameters on top of those given by Eqs.~\eqref{uppervev} and~\eqref{uppervevtwo}. Particles $\rho$ typically decay into axions at some point during cosmological evolution, at least for not very large $f^{0}_{PQ}$. 
This is a perturbative decay, which is quite slow and thus occurs rather late. Produced axions are ultra-relativistic and therefore contribute to dark radiation. Existing cosmological constraints on dark radiation 
lead to strong bounds on the model constants, as we discuss in details in Subsection~\ref{sec:perturbative}. The bounds are slightly relaxed, if $\rho$-particles thermalize with the primordial plasma (Subsection~\ref{sec:thermal}), which is generically the case for astrophysically interesting 
values $f^{0}_{PQ} \sim 10^{8}~\mbox{GeV}$.

\subsection{Perturbative decay of $\rho$-particles}
\label{sec:perturbative}

In this subsection, we assume that $\rho$-particles are disconnected from thermal bath, so that their number density is approximately conserved in the comoving volume until the moment, when they decay into a couple of relativistic axions. 
The decay rate is given by
\begin{equation}
\label{decayrate}
\Gamma_{dec} = \frac{(M^0_S)^3 }{32\pi (f^{0}_{PQ})^2} = \frac{v^6_{\phi}}{32\pi \cdot (2\beta)^{3/2} \cdot (f^{0}_{PQ})^5} \; ,
\end{equation}
where we used Eq.~\eqref{boundtwo} in the second equality. The  temperature at the time of the decay, $T_{dec}$, is estimated from $\Gamma_{dec} \simeq H(T_{dec})$. Using the value~\eqref{betaaxionDM} of the constant $\beta$ defined by the requirement 
that axions make most of the dark matter, we obtain
\begin{equation}
\label{decaytemp}
T_{dec} \simeq 30~\mbox{keV} \cdot \left(\frac{4}{N} \right)^{9/4} \cdot \left(\frac{g_* (T_i)}{100} \right)^{3/2} \cdot \left(\frac{100}{g_* (T_{dec})} \right)^{1/4}  \cdot \left(\frac{v_{\phi}}{1~\mbox{TeV}} \right)^{9/2} \cdot \left(\frac{10^8~\mbox{GeV}}{f^{0}_{PQ}} \right)^{5/2} \; .
\end{equation}
Another important quantity in our following discussions is the temperature of the Universe $T_*$ at the moment when the energy density ${\cal E}_{\rho}$ of the field $\rho$ (formally assumed to be stable) becomes equal to that of radiation. 
The energy density ${\cal E}_{\rho} (t)$ is estimated as
\begin{equation}
\label{energyrho}
{\cal E}_{\rho} (t>t_{PT}) \simeq \frac{r \lambda_S f^4_{PQ,i}}{4} \cdot \left(\frac{R_i}{R_{PT}} \right)^4 \cdot \left(\frac{R_{PT}}{R(t)} \right)^3 \; .
\end{equation}
The subscript `$PT$' stands for a phase transition, which occurs at the temperature $T_{PT} \simeq v_{\phi}$. In Eq.~\eqref{energyrho}, we have taken into account that the field $\rho$ including its fluctuations, evolve as radiation before the phase transition, and later on as non-relativistic 
fluid, because the mass $M_S$ relaxes to a constant value at $T_{PT}$ simultaneously with the scale $f_{PQ}$\footnote{This is true provided that $k/a_{PT} \lesssim M^{0}_S$, where $k$ are characteristic momenta of the field $\rho$ 
fluctuations produced in the narrow parametric resonance regime. This inequality can be rephrased as $k/(a_i M_S (t_i)) \lesssim \sqrt{12/N} \cdot \left(g_* (T_i)/g_* (T_{PT}) \right)^{1/3}$. In what follows, we assume that momenta of produced particles $\rho$ 
are of the order of $M_S (t_i)$, so that the latter inequality is fulfilled.}. There we also neglected the contribution following from the freeze-in production of $\rho$-particles by scattering of thermal 
particles $\phi$. This is justified for the tiny constants $g$ as in Eq.~\eqref{g} and small mass values~\eqref{masslow}~\cite{Lebedev:2019ton, Chu:2011be}.  Using Eqs.~\eqref{beta},~\eqref{vevhot}, and entropy conservation in the comoving volume, we rewrite Eq.~\eqref{energyrho} as
\begin{equation}
{\cal E}_{\rho} (t>t_{PT}) \simeq \frac{7 \cdot 10^{-5}}{\beta} \cdot \left(\frac{N}{4} \right)^2  \cdot \left(\frac{r}{0.01} \right) \cdot \frac{g^{1/3}_* (T_{PT}) g_* (T)}{g^{4/3}_* (T_i)} \cdot v_{\phi} \cdot T^3(t) \; .
\end{equation}
Substituting the value~\eqref{betaaxionDM}, we obtain 
\begin{equation}
\label{equality}
T_* \simeq 60~\mbox{keV} \cdot \left(\frac{4}{N} \right) \cdot \left(\frac{g_* (T_i)}{100} \right)^{2/3} \cdot \left(\frac{g_* (T_{PT})}{100} \right)^{1/3} \cdot \left(\frac{r}{0.01} \right) \cdot \left(\frac{v_{\phi}}{1~\mbox{TeV}} \right)^3  \; .
\end{equation}
Note that Eq.~\eqref{betaaxionDM} and hence Eqs.~\eqref{decaytemp} and~\eqref{equality} must be revisited, if axions give a negligible contribution to dark matter -- the option discussed in the end of this section. 

Depending on the decay rate of particles $\rho$, their cosmological applications can be separated into three distinct scenarios. 

\begin{figure}[tb!]
  \begin{center}
    \includegraphics[width=0.75\columnwidth,angle=-90]{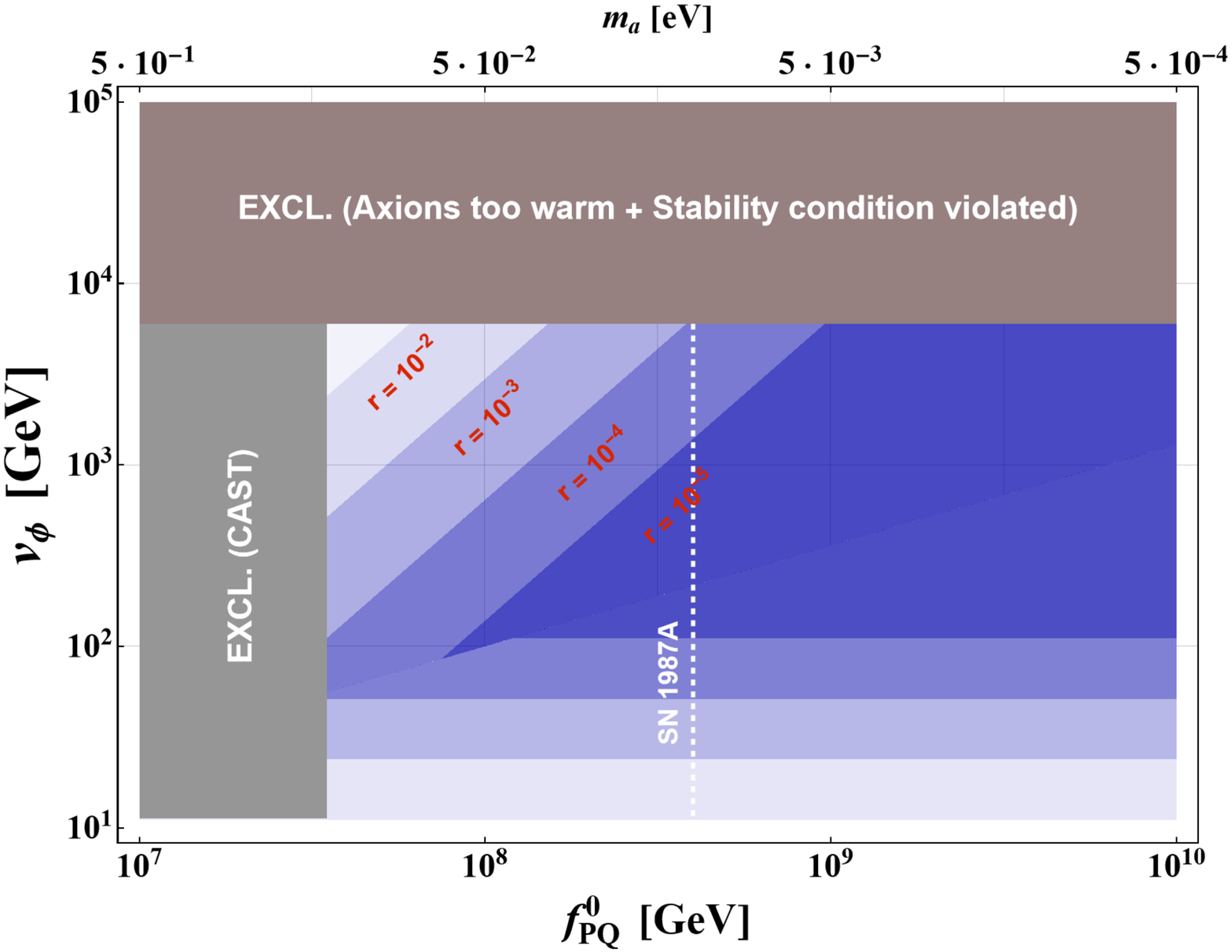}
  \caption{Allowed model parameter space is shown for the case when axions make most of the dark matter in the Universe, in terms of the thermal field $\phi$ expectation value $v_{\phi}$ and low-temperature Peccei-Quinn symmetry breaking scale $f^{0}_{PQ}$ 
  for different values of the ratio $r$ defined by Eq.~\eqref{ratio}. We have set $g_* (T_i)=g_*(T_{dec})=g_*(T_{PT})=100$, $\lambda_{\phi}=1$, and $N=4$. It is assumed that particles $\rho$ are not thermalized with primordial plasma. The part of parameter space excluded through the impact of decaying $\rho$-particles 
  on the CMB, grows and moves from the dark blue region towards the light blue one, as the ratio $r$ is increased. Tentative bound following from the impact of axion emission on the duration of neutrino burst from SN 1987A is shown 
  with the dashed line.
  }\label{parameterspace}
  \end{center}
\end{figure}

{\it Particles $\rho$ decay before matter-radiation equality: $T_{dec} \gtrsim T_{eq}$.}  This is perhaps the best motivated option. Indeed, using Eq.~\eqref{decaytemp}, we can convert the condition $T_{dec} \gtrsim T_{eq}$ into the limit 
\begin{equation}
\label{rangeone}
v_{\phi} \gtrsim 100~\mbox{GeV} \cdot \left(\frac{N}{4} \right)^{1/2} \cdot \left(\frac{100}{g_* (T_i)} \right)^{1/3}  \cdot \left(\frac{g_* (T_{dec})}{100} \right)^{1/18} \cdot \left(\frac{f^{0}_{PQ}}{10^{8}~\mbox{GeV}} \right)^{5/9} \; ,
\end{equation}
which covers the case of the Higgs field and not very large Peccei--Quinn symmetry breaking scale. This scenario is constrained by the requirement that decay products of the field $\rho$, -- axions -- 
do not affect the BBN and the CMB. The fraction of dark radiation composed of these ultra-relativistic axions is conventionally expressed in terms of the effective number of neutrino species $N_{eff}$. Our model predicts 
the following departure $\Delta N_{eff} \equiv N_{eff} -N_{eff, SM}$ from the Standard Model value $N_{eff, SM} \approx 3.046$: 
\begin{equation}
\label{deltaneff}
\Delta N_{eff} \approx \frac{4}{7}  \cdot \frac{g_* (T_{dec}) \cdot T_*}{T_{dec}} \cdot \left[\frac{43}{4 g_* (T_{dec})} \right]^{4/3} \ .
\end{equation}
Note that the factor $\sim g_* (T_{dec}) \cdot T_*/T_{dec}$ corresponds to the number of degrees of freedom deployed in primordial plasma at the time of the decay. The factor in the square brackets takes into account that the axions never get thermalized with primordial plasma, which is heated by the annihilations of particles. 
Combination of Planck and BAO data fixes the value $N_{eff} =2.99 \pm 0.17$ at $68\%$ CL~\cite{Planck:2018vyg}, so that $\Delta N_{eff} \lesssim 0.11$. From this limit, using Eqs.~\eqref{decaytemp},~\eqref{equality}, and Eq.~\eqref{deltaneff}, we obtain
\begin{equation}
\begin{split}
v_{\phi} &\gtrsim 14~\mbox{TeV}  \cdot \left(\frac{N}{4} \right)^{5/6} \cdot  \left(\frac{100}{g_* (T_{dec})} \right)^{1/18}  \cdot \left(\frac{100}{g_* (T_i)} \right)^{5/9} \cdot \left(\frac{g_* (T_{PT})}{100} \right)^{2/9} \times \\ & \times  \left(\frac{r}{0.01} \right)^{2/3} \cdot \left(\frac{f^{0}_{PQ}}{10^{8}~\mbox{GeV}} \right)^{5/3} \; . 
\end{split}
\end{equation}
Avoiding the conflict with the limits~\eqref{uppervev} and~\eqref{uppervevtwo}, we should require that the energy density of particles $\rho$ is small relative to the axion energy density by the end of parametric resonance. In particular, 
if it turns out that $r \simeq 0.01$, the parameter space is quite fixed: $v_{\phi} \simeq 10~\mbox{TeV}$ and $f^{0}_{PQ} \simeq 10^{8}~\mbox{GeV}$, see Fig.~\ref{parameterspace}. For $r \gtrsim 0.01$, one should assume 
that $f^{0}_{PQ} \lesssim 10^{8}~\mbox{GeV}$. While allowed by the current CAST limit on the axion-photon coupling\footnote{For example, in KSVZ scenario axion-photon coupling is related to Peccei--Quinn scale $f^{0}_{PQ}$ by 
$g_{a\gamma \gamma} \approx \alpha/(\pi f^{0}_{PQ})$, where $\alpha \approx 1/137$ is the fine-structure constant. Consequently, the CAST bound on $g_{a \gamma \gamma}$ implies the limit $f^{0}_{PQ} \gtrsim 3.5 \cdot 10^{7}$~\mbox{GeV} at $95\%$ CL.}~$g_{a\gamma \gamma} \lesssim 0.66 \times ~10^{-10}~\mbox{GeV}^{-1}$ at $95\%$ CL~\cite{CAST:2017uph}, 
they are in a gross conflict with the supernova 1987A rough bound $f_{PQ} \gtrsim 4 \times 10^{8}~\mbox{GeV}$. If the latter is confirmed in the future and 
if lattice simulations of parametric resonance show large ratios $r \gtrsim 0.01$, our model of axion dark matter production will be jeopardized in the range~\eqref{rangeone}. 
This conclusion is avoided in the important situation that $\rho$-particles are thermalized with primordial bath discussed in the next subsection.

{\it Particles $\rho$ decay after matter-radiation equality: $T_0 \lesssim T_{dec} \lesssim T_{eq}$.} This option assumes the following range of expectation values $v_{\phi}$: 
\begin{equation}
v_{\phi} \simeq [8-100]~\mbox{GeV}  \cdot \left(\frac{N}{4} \right)^{1/2} \cdot \left(\frac{100}{g_* (T_i)} \right)^{1/3} \cdot \left(\frac{g_* (T_{dec})}{100} \right)^{1/18} \cdot \left(\frac{f^{0}_{PQ}}{10^{8}~\mbox{GeV}} \right)^{5/9} \; .
\end{equation}
 In that case, the field $\rho$ constitutes an unstable fraction of dark matter estimated as $\sim T_*/T_{eq}$. The unstable component should not contribute more than $10\%$ to the overall dark matter energy density~\cite{Berezhiani:2015yta}, 
so that $T_*/T_{eq} \lesssim 0.1$. Strictly speaking, this bound assumes dark matter decaying after recombination, but we ignore this subtlety given a small separation in time between matter-radiation equality and recombination. By virtue of Eq.~\eqref{equality}, we obtain the limit on the ratio $r$:
\begin{equation}
\label{extreme}
r \lesssim 5 \cdot 10^{-4} \cdot \left(\frac{N}{4} \right) \cdot \left(\frac{100}{g_* (T_i)} \right)^{2/3} \cdot \left(\frac{100}{g_* (T_{PT})} \right)^{1/3} \cdot \left(\frac{30~\mbox{GeV}}{v_{\phi}} \right)^3\; .
\end{equation}
The upper bound here appears to be quite stringent and assumes that production of particles $\rho$ during parametric resonance is extremely inefficient. Furthermore, small expectation values $v_{\phi}$ assumed in this scenario 
can be problematic from the viewpoint of embedding it into KSVZ or DFSZ framework. The bound~\eqref{extreme} can be relaxed considerably, if we abandon 
the restriction that axions make all of the dark matter. In that case, however, more interesting is the following scenario.

{\it Particles $\rho$ are stable throughout cosmological history: $T_{dec} \lesssim T_0$.} If this inequality is fulfilled, one can consider the particles $\rho$ as stable dark matter candidates. There are two interesting sub-scenarios. First, let us assume that axions 
still constitute most of dark matter in the Universe, while the particles $\rho$ contribute a small fraction $\xi$, i.e., $T_* \simeq \xi T_{eq}$. Using Eq.~\eqref{equality}, we obtain the constraint on the model parameters for a given $\xi$:
\begin{equation}
\label{toosmall}
v_{\phi} \simeq 20~\mbox{GeV} \cdot \xi^{1/3}  \cdot \left(\frac{N}{4}\right)^{1/3} \cdot \left(\frac{100}{g_* (T_i)} \right)^{2/9} \cdot \left(\frac{100}{g_* (T_{PT})} \right)^{1/9} \cdot \left(\frac{0.01}{r} \right)^{1/3}\; .
\end{equation}
On the other hand, from the stability of particles $\rho$ on cosmological time scales, one gets
\begin{equation}
 v_{\phi} \ll 8~\mbox{GeV} \cdot \left(\frac{N}{4} \right)^{1/2} \cdot \left(\frac{100}{g_* (T_i)} \right)^{1/3} \cdot \left(\frac{g_* (T_{dec})}{100} \right)^{1/18}  \cdot \left(\frac{f^{0}_{PQ}}{10^{8}~\mbox{GeV}} \right)^{5/9} \; .
\end{equation}
Notably, this scenario works well also for large values of $r \simeq 1$ and $f^{0}_{PQ} \gg 10^{8}~\mbox{GeV}$. Nevertheless, as in the case of unstable dark matter considered earlier, for such small  values of $v_{\phi}$ suggested in Eq.~\eqref{toosmall}, the possibility of embedding into KSVZ or DFSZ type of models is questionable. This explains why the range of values of $v_{\phi}$ is cut at $10~\mbox{GeV}$ in Fig.~\ref{parameterspace}.

Finally, we consider the sub-scenario, where most of dark matter is constituted by the particles $\rho$. In this case, the constant $\beta$ is given by
\begin{equation}
\label{betarho}
\beta \simeq 2.5 \cdot 10^6  \cdot \left(\frac{N}{4} \right)^2 \cdot \left(\frac{100}{g_* (T_i)} \right)^{4/3}  \cdot \left(\frac{g_* (T_{PT})}{100} \right)^{1/3} \cdot \left(\frac{r}{0.01} \right)\cdot \left(\frac{v_{\phi}}{1~\mbox{TeV}} \right) \; ,
\end{equation}
which is obtained by equating the energy density~\eqref{energyrho} to radiation energy density at the time $t_{eq}$. For so large values of the constant $\beta$, axions give a negligible contribution to dark matter, as Eq.~\eqref{betaaxionDM} is grossly violated. One should require that the lifetime of the field $\rho$ exceeds the age of the Universe, i.e., $\Gamma_{dec } \gtrsim H_0$. Substituting Eq.~\eqref{betarho} into Eq.~\eqref{decayrate}, we obtain
\begin{equation}
f^{0}_{PQ} \gtrsim 3.5 \cdot 10^{9}~\mbox{GeV}  \cdot \left(\frac{4}{N} \right)^{3/5} \cdot \left(\frac{g_* (T_i)}{100} \right)^{2/5} \cdot \left(\frac{100}{g_* (T_{PT})} \right)^{1/10} \cdot \left(\frac{0.01}{r} \right)^{3/10} \cdot \left(\frac{v_{\phi}}{1~\mbox{TeV}} \right)^{9/10} \; .
\end{equation}
Comparing with the values $f^{0}_{PQ} \simeq 10^{11}-10^{12}~\mbox{GeV}$ characteristic for the misalignment mechanism, we observe that there is a rather broad window in the parameter space, where the field $\rho$ can play the role of dark matter. Note that for large values of $\beta$ as in Eq.~\eqref{betarho}, the radial field $M^{0}_S$ is rather light, see Eq.~\eqref{boundtwo}.

 \subsection{Thermalization of $\rho$-particles}
 \label{sec:thermal}

We get back to the scenario, where axions make most of dark matter. As it has been mentioned above, for astrophysically most interesting values $f^{0}_{PQ} \sim 10^{8}~\mbox{GeV}$, the 
particles $\rho$ may thermalize with the plasma. This can be achieved in the KVSZ scenario involving additional heavy quarks~\cite{Kim:1979if, Shifman:1979if}. Integrating out heavy quarks, 
one ends up with an effective interaction between the field $S$ and gluons, which leads to thermalization of $\rho$-particles with the rate~\cite{Mukaida:2012qn, Graf:2012hb, Moroi:2014mqa}
\begin{equation}
\label{dissipation}
\Gamma_{th} \simeq \frac{\alpha^2_{QCD}}{16 \pi^2 \ln \alpha^{-1}_{QCD}} \cdot \frac{T^3}{f^2_{PQ}} \; .
\end{equation}
Efficiency of thermalization is defined by the ratio $\Gamma_{th}/H$. The latter grows as $1/T$ at temperatures $T \gtrsim v_{\phi}$, and later on decreases as $T$. Consequently, the ratio 
$\Gamma_{th}/H$ is maximal at the temperature $T \simeq v_{\phi}$, and it can be expressed as 
\begin{equation}
\label{atmax}
\left(\frac{\Gamma_{th}}{H} \right)_{max} \simeq  \left(\frac{16}{N} \right) \cdot \sqrt{\frac{100}{g_* (T_{PT})}} \cdot \left(\frac{v_{\phi}}{1~\mbox{TeV}} \right) \cdot \left(\frac{10^{8}~\mbox{GeV}}{f^{0}_{PQ}} \right)^2 \; ,
\end{equation} 
where we used $\alpha_{QCD} \approx 0.1$. We observe that thermalization taking place for $\left( \Gamma_{th}/H \right)_{max} \gtrsim 1$ bounds the Peccei-Quinn scale to the range $f^{0}_{PQ} \simeq (1\text{--}5) \times 10^{8}~\mbox{GeV}$ 
for high enough expectation values $v_{\phi} \simeq 1\text{--}10~\mbox{TeV}$. At the same time, thermalization by this mechanism in the case of the SM Higgs field remains 
questionable and deserves a more elaborate investigation, which we leave out of the scope of this work.

The rate of axion thermal production is also estimated using Eq.~\eqref{dissipation}~\cite{Masso:2002np, Graf:2010tv, Salvio:2013iaa}. It is crucial, however, that axions generated during the parametric resonance do not dissipate in the primordial plasma. 
As it is discussed in Ref.~\cite{Moroi:2014mqa}, the rate of their dissipation is suppressed compared to Eq.~\eqref{dissipation} due to low 
axion momenta $|{\bf p}| \sim M_{S}$, i.e., by the factor $\sim M^2_S/(g^4_{QCD} T^2)$. Given that $M_S \sim g T$, we estimate the rate of axion dissipation as 
\begin{equation}
\Gamma^{a}_{diss} \sim \left(\frac{g}{g^2_{QCD}} \right)^2 \cdot \Gamma_{th} \; .
\end{equation}
It is evident from Eqs.~\eqref{g} and~\eqref{atmax} that $\Gamma^{a}_{diss} \ll H$, i.e., the comoving number density of axions produced through the parametric resonance, remains intact. 
Other processes involving axions also do not lead to their dissipation in the plasma. In particular, the dissipation rate due to the process $a+\phi \rightarrow a +\phi$, which involves 
a virtual $\rho$-particle, is estimated as $\sim g^4 T^3/(8\pi^3 m^2_{\phi})$, where $m_{\phi}$ is the thermal mass of the field $\phi$. With minuscule values of $g \lesssim 10^{-4}$ assumed, this process is indeed very 
slow. The processes $a+\rho_{th} \rightarrow a_{th}+\rho_{th}$ and $a+a_{th} \rightarrow a_{th}+a_{th}$ involving interactions with thermal axions and $\rho$-particles, are further suppressed.

Thermally produced axions make a negligible contribution to the effective number of neutrino species, i.e., $\Delta N_{eff} \approx 0.027$~\cite{Planck:2018vyg}, because they decouple early from the primordial plasma.  
This would be also the case of $\rho$-particles, if they remained relativistic down to the moment of their decay into axions at the temperature $T_{dec}$. However, 
for viable values of parameters $v_{\phi} \lesssim 10~\mbox{TeV}$, see Eqs.~\eqref{uppervev} and~\eqref{uppervevtwo}, and $f^{0}_{PQ} \gtrsim 10^{8}~\mbox{GeV}$, one has $M^{0}_S>T_{dec}$. Consequently, the energy density of $\rho$-particles grows relative to that of radiation 
for the temperatures $T_{dec} \lesssim T \lesssim M^{0}_S$. This leads to larger $\Delta N_{eff}$ compared to naive $\sim 0.03$ by the factor $\sim M^{0}_S/T_{dec}$. Using Eqs.~\eqref{masslow} and~\eqref{decaytemp}, we obtain  
\begin{equation}
\Delta N_{eff} \simeq 0.03 \cdot \left(\frac{N}{4} \right)^{3/4} \cdot \sqrt{\frac{100}{g_* (T_i)}} \cdot \left(\frac{g_* (T_{dec})}{100} \right)^{1/4} \cdot \left(\frac{10~\mbox{TeV}}{v_{\phi}} \right)^{3/2} \cdot \left(\frac{f^{0}_{PQ}}{10^{8}~\mbox{GeV}}  \right)^{3/2} \; . 
\end{equation}
As in the previous section, the Planck+BAO $68\%$ CL limit $\Delta N_{eff} \lesssim 0.11$~\cite{Planck:2018vyg} constrains the model parameter space in the narrow range close to $v_{\phi} \sim 10~\mbox{TeV}$ and $f^{0}_{PQ} \sim 10^{8}~\mbox{GeV}$. 
However, now this constraint is applicable independently of the energy density of $\rho$-particles produced during the prametric resonance, meaning that the parameter $r$ can be as large as $r \sim 1$.

\section{Discussions} 
\label{sec:Dis}

In this work, we considered an axion scenario involving the temperature-dependent scale of Peccei--Quinn symmetry breaking $f_{PQ}$. The temperature dependence is introduced through the simple portal interaction of the complex Peccei-Quinn field $S$ and a thermal Higgs-like scalar $\phi$ with a non-zero expectation value $v_{\phi}$. As the Universe's temperature drops below $v_{\phi}$, 
the scale $f_{PQ}$ relaxes to a constant value $f^{0}_{PQ}$. In this scenario, axion particles can be naturally created by coherent oscillations of the radial field $\rho \equiv \sqrt{2}|S|$ with the amplitude $\sim f_{PQ} (t)$. Notably, axion production turns out to be very efficient independently of the Peccei-Quinn symmetry breaking scale, see Figs.~\ref{narrowresonance} and~\ref{resonance}. We have shown in Section~\ref{sec:darkmatter} that axions can have the right abundance and their momenta are sufficiently small to explain the cold dark matter in the Universe. In particular, this is fulfilled for the values of $f^{0}_{PQ}$ well below those required by the misalignment mechanism.

A more detailed investigation of the parametric resonance with lattice simulations is required. The crucial task to be accomplished is obtaining the ratio $r$ defined by Eq.~\eqref{ratio}, i.e., the fraction of $\rho$-particles in the total 
energy density carried by the field $S$. 
Even for a tiny fraction $r$, particles $\rho$ may strongly impact BBN and CMB. On the other hand, if radial excitations efficiently thermalize with the plasma, the conflict with cosmological observations can be avoided independently of 
$r$ value, 
which is indeed the case for $f^{0}_{PQ} \sim 10^{8}~\mbox{GeV}$ and $v_{\phi} \sim 10~\mbox{TeV}$.
At the moment, we deem this to be vanilla spot of parameter space. In particular, it remains an open issue, if the scenario with the portal coupling to the SM Higgs field is viable. This scenario requires $r \lesssim 10^{-4}$, see Fig.~\ref{parameterspace}; it is also 
in tension with evolution of red giants according to the discussion in the end of Section~\ref{sec:darkmatter}. However, with future lattice simulations and more detailed analysis of the particles $\rho$ thermalization 
in the plasma the range of model parameters can be reassessed.

Another important issue is embedding our Peccei--Quinn symmetry breaking mechanism into a proper particle physics framework, KSVZ or DSFZ. 
This can be accomplished straightforwardly, at least for large expectation values $v_{\phi} \gtrsim 100~\mbox{GeV}$. Namely, one simply replaces the bare tachyonic mass of the field $S$ in these scenarios 
by the portal interaction with the scalar $\phi$. The choice of DFSZ type of models is particularly well motivated (apart from the aforementioned tensions), as they naturally assume portal couplings of the field $S$ to Higgs doublets.

 Our scenario can be probed with observational data in multiple ways. i)~As the small scales of Peccei--Quinn symmetry breaking are concerned, we anticipate interesting applications 
 for supernovae and horizontal branch stars. At the moment, the often cited supernova limit $f_{PQ} \gtrsim 4 \times 10^{8}~\mbox{GeV}$ is plagued by various uncertainties, cf. Ref.~\cite{Bar:2019ifz}. These uncertainties can be resolved with the future supernovae observations~\cite{Fischer:2016cyd}. ii) Planned helioscopes and haloscopes, e.g., IAXO~\cite{IAXO:2019mpb} and TASTE~\cite{TASTE:2017pdv} will have a potential to improve the bound on the axion-photon coupling compared to the 
 present CAST limit~\cite{CAST:2017uph}. iii) In the vanilla part of parameter space corresponding to $v_{\phi} \simeq 10~\mbox{TeV}$, axions can be relatively warm, see Eq.~\eqref{uppervevtwo}. 
 Consequently, one may expect a non-trivial imprint on Lyman $\alpha$ forest from distant quasars. Note that this situation is rather generic for axions produced through the parametric resonance~\cite{Co:2017mop, Harigaya:2019qnl, Moroi:2020has}. iv) According to the discussion in Section~\ref{sec:cosmology}, particles $\rho$ may impact the BBN and the CMB through their decay products -- relativistic axions, which contribute to dark radiation. Particles $\rho$ 
can also make up a small (unstable) fraction or even entire dark matter density. v) There is a possibility of stochastic gravitational wave emission caused by axion perturbations amplified during the parametric resonance~\cite{Figueroa:2017vfa}. Precise determination of the peak amplitude and spectrum of gravitational waves is out of the scope of the present work. Nevertheless, one can make a rough estimate of their peak frequency at production $f_{gw} \sim H_i$. The present-day peak frequency is estimated as~\cite{Ramazanov:2021eya}
\begin{equation}
f_{gw} (t_0) \sim 60~\mbox{Hz} \cdot \sqrt{N}  \cdot \left(\frac{100}{g_* (T_i)} \right)^{1/3} \cdot \left(\frac{g}{10^{-8}} \right)\; . 
\end{equation}
Thus, for the constants $g \lesssim 10^{-8}$, gravitational waves are in the frequency range of Einstein Telescope, DECIGO, or LISA experiments. Gravitational waves can  also be produced at a phase transition taking place at the temperature 
$T \sim v_{\phi}$. The resulting peak frequencies again fall in the observable range for the expectation values of interest $100~\mbox{GeV} \lesssim v_{\phi} \lesssim 10~\mbox{TeV}$, cf. Ref.~\cite{Caprini:2019egz}.

\section*{Acknowledgments}
We are indebted to Dmitry Gorbunov, Alexander Vikman, and Wen Yin for useful discussions. The work of S.~R. and R.~S. is supported by the Czech Science Foundation, GA\v CR, grant number 20-16531Y, European
Structural and Investment Funds (ESIF/ERDF) and the Czech Ministry of Education, Youth and Sports (M\v SMT) through the Project CoGraDS -CZ.02.1.01/0.0/0.0/15003/0000437. 
R.~S. also acknowledges the project MSCA-IF IV FZU - CZ.02.2.69/0.0/0.0/20 079/0017754, European Structural and Investment Fund, and the Czech Ministry of Education, Youth and Sports.

\end{document}